\documentclass[preprint]{aastex63}

\usepackage{graphicx}
\usepackage{xspace}
\usepackage{bm}

\usepackage{xcolor}

\newcommand{\code}[1]{\texttt{#1}}

\newcommand{\pwv}{\ensuremath{\rm PWV}\xspace}
\newcommand{\pwvzenith}{\ensuremath{{\rm PWV}_{\rm zenith}}\xspace}
\newcommand{\pwvwade}{\ensuremath{{\rm PWV}_{\rm Wade}}\xspace}
\newcommand{\pwveff}{\ensuremath{{\rm PWV}_{\rm los}^{\rm eff}}\xspace}
\newcommand{\pwvlos}{\ensuremath{{\rm PWV}_{\rm los}}\xspace}
\newcommand{\deltazbluered}{\ensuremath{\Delta z({\rm blue} - {\rm red})}\xspace}

\graphicspath{{figures/}}

\begin{document}

\title{GPS Measurements of Precipitable Water Vapor Can Improve Survey Calibration: A Demonstration from KPNO and the Mayall z-band Legacy Survey}

\author{W.~M.~Wood-Vasey}
\affiliation{
	Pittsburgh Particle Physics, Astrophysics, and Cosmology Center (PITT PACC).
	Physics and Astronomy Department, University of Pittsburgh,
	Pittsburgh, PA 15260, USA
}

\author{Daniel Perrefort}
\affiliation{
	Pittsburgh Particle Physics, Astrophysics, and Cosmology Center (PITT PACC).
	Physics and Astronomy Department, University of Pittsburgh,
	Pittsburgh, PA 15260, USA
}

\author{Ashley D. Baker}
\affiliation{
    University of Pennsylvania Department of Physics and Astronomy.
    209 S 33rd Street, Philadelphia, PA 19104, USA
}
\affiliation{
    Department of Astronomy, California Institute of Technology, Pasadena, CA 91125, USA
}

\begin{abstract}
Accounting for the effects of atmospheric absorption is a key step in calibrating high-accuracy photometry.
In particular, the atmospheric transmission redward of 550~nm is dominated by absorption due to water vapor.
We here show that dual-band GPS measurements of precipitable water vapor (PWV) at the Kitt Peak National Observatory (KPNO) predict the overall per-image sensitivity of the Mayall $z$-band Legacy Survey (MzLS).
We further find that the per-image variation in the brightness of individual stars is strongly correlated with the measured PWV and the color of the star.
We use synthetic stellar spectra and TAPAS transmission models to predict the expected PWV-induced photometric errors and find good agreement with the observations.
In line with previous work in the literature we also find that PWV absorption can be well-approximated by a linear relationship with \pwveff and present an update on the traditional treatment in the literature.

Within the range of reasonable observing conditions, the MzLS zero point varies with a standard deviation of 127~mmag.  This variation is dominated by a gray secular trend with time, consistent with a gradual accumulation of contamination on optical surfaces that accounts for $\sim$114~mmag of variation.
Correcting for PWV based on a suite of stellar spectra and detailed PWV absorption models accounts for another 47~mmag of zero-point variation.
The MzLS per-image sensitivity is decreased by $\sim$40~mmag per effective mm of PWV.
The difference between blue ($r-z < 0.5$~mag) and red ($1.2$~mag~$< r-z$) stars
increases by 3.25 mmag per effective mm of PWV.

These results show the need for high-precision photometric surveys to simultaneously monitor PWV.
We find that GPS systems provide more precise PWV measurements than using differential measurements of stars of different colors and recommend that observatories that observe long-ward of 800~nm install dual-band GPS as a low-maintenance, relatively low cost, auxiliary calibration system.
We extend our results of the need for well-calibrated PWV measurements by presenting calculations of the PWV photometric impact on three science cases of interest: stellar photometry, supernova cosmology, and quasar identification and variability.

\end{abstract}

\section{Introduction}

Modern astronomical surveys such as the Dark Energy Survey \citep[DES;][]{DES2016} and the Panoramic Survey Telescope and Rapid Response System \citep[Pan-STARRS;][]{Magnier16} require high photometric accuracy, and upcoming surveys such as the Vera Rubin Observatory Legacy Survey of Space and Time \citep[LSST;][]{Ivezic2019} will strive to push this limit even further by requiring a 5~mmag photometric accuracy.
For surveys relying on ground-based telescopes, accounting and correcting for effects induced by the Earth's atmosphere is a key step in achieving this goal \citep[see][]{Ivezic07, Magnier16, Burke10, Burke14, Burke18}.
In particular, absorption features due to the presence of water vapor dominate the atmospheric transmission function in the red optical and near-infrared (NIR) band-passes ($\lambda > 550$~nm).
The complicated yet known structure of water vapor absorption must be accurately and precisely accounted for to accurately determine the complete system throughput as a function of wavelength.

Broad-band imaging is traditionally calibrated using a reference catalog to compute correction terms for color, airmass, and a higher-order color-airmass term.
While this approach implicitly corrects for first-order effects introduced by the atmospheric opacity, it does not account for second-order effects caused by differences in the atmospheric absorption between the spectral energy distributions of the target and reference stars.
Because redder stars emit much more of their light at wavelengths susceptible to atmospheric absorption, their photometric values vary differently than bluer stars with changes in atmospheric conditions. This difference in behavior can introduce second-order photometric errors of over 1\% \citep{Ivezic07, Li16}, which is significant in an era of high-accuracy photometry.

Approaches that allow for time-dependent color terms can account for time-variable absorption.
However, these color terms work best where the differences in the SEDs can be described by smooth monotonic functions, such as in the Rayleigh-Jeans tail of the effective blackbody approximation for K-type and hotter stars.
This approach works particularly well for effects that themselves are smooth in wavelength, such as Mie scattering due to atmospheric aerosols.
However, the contribution to the atmospheric absorption due to atmospheric water vapor is not smooth.

Unlike the attenuation of light due to ozone and aerosols, absorption due to precipitable water vapor (PWV) has a complex transmission function with lines that saturate even during relatively dry and \textit{photometric} conditions (PWV $\approx 5$ mm).
The convention is to measure water vapor in units of the mm of liquid water if it were all condensed.
Local PWV concentrations can change by up to $10\%$ per hour \citep{Li17}, requiring a photometric correction that varies over the course of an observing night.
Correctly accounting for variable complex atmospheric transmission requires a detailed understanding of the atmospheric state during the time in which observations were performed.

A common astronomical method for quantifying the absorption due to the atmosphere is to perform dedicated observations of a young A-type star \citep[e.g., as in][]{Stubbs07}.
Because these stars have relatively few intrinsic features and are well described by existing models, the effects of atmospheric absorption can be fit for using forward modeling.
When performed spectroscopically, fitting these observations provides a detailed map of the per-wavelength transmission of the atmosphere.
This approach can also be performed photometrically by using narrow-band filters centered on key atmospheric features \citep{Li14,Baker17}.
This narrow-band image approach benefits from the ability to perform in situ measurements along a similar line of sight to the scientific observations. Both of these approaches require the maintenance and calibration of a secondary, dedicated telescope.

An alternative method to measure the water absorption is to use dual-band receivers tracking signals from Global Positioning Satellites (GPS). By measuring the delay of dual-band GPS signals traveling through the atmosphere, it is possible to determine the PWV column density along zenith \citep{Nahmias04, Blake11, Manandhar18}. Through the use of atmospheric models and scaling relations, the atmospheric transmission due to PWV can be determined for the line-of-sight airmass of a given observation. This approach benefits from the ability to perform atmospheric measurements in near real-time and results in values that correlate strongly with spectroscopic measurements \citep{Li17}. However, GPS measurements tend to be less accurate in dry conditions \citep{Buehler12, Hagemann03} and cannot constrain the transmission due to secondary (non-PWV) components of the atmosphere.

In this work, we demonstrate the effect of PWV on the measured zero points and color-dependent per-image offsets
using observations taken by the Mayall $z$-band Legacy Survey (MzLS)
combined with contemporaneous GPS PWV measurements from Kitt Peak National Observatory (KPNO).

In Section \ref{sec:data} we outline the MzLS observations (\ref{sec:mzls_data}), PWV measurements (\ref{sec:gps_pwv}), and atmospheric models (\ref{sec:telluric_models}) considered by this work.
In Section \ref{sec:mzls_results} we demonstrate the effects of PWV absorption on MzLS zero points
and compare using GPS-measured PWV to using the differential brightness of stars of different colors to predict zero point variation.
We then consider the impact of PWV on various science cases in Section \ref{sec:science_cases}, including the impact on stars (\ref{sec:stars}), supernovae (\ref{sec:supernovae}), and quasars (\ref{sec:quasars}).
Finally, we discuss the application of PWV measurement techniques to upcoming surveys in Section~\ref{sec:conclusions}.

\section{Data} \label{sec:data}

\subsection{Mayall $z$-band Legacy Survey} \label{sec:mzls_data}

\begin{figure*}
    \plotone{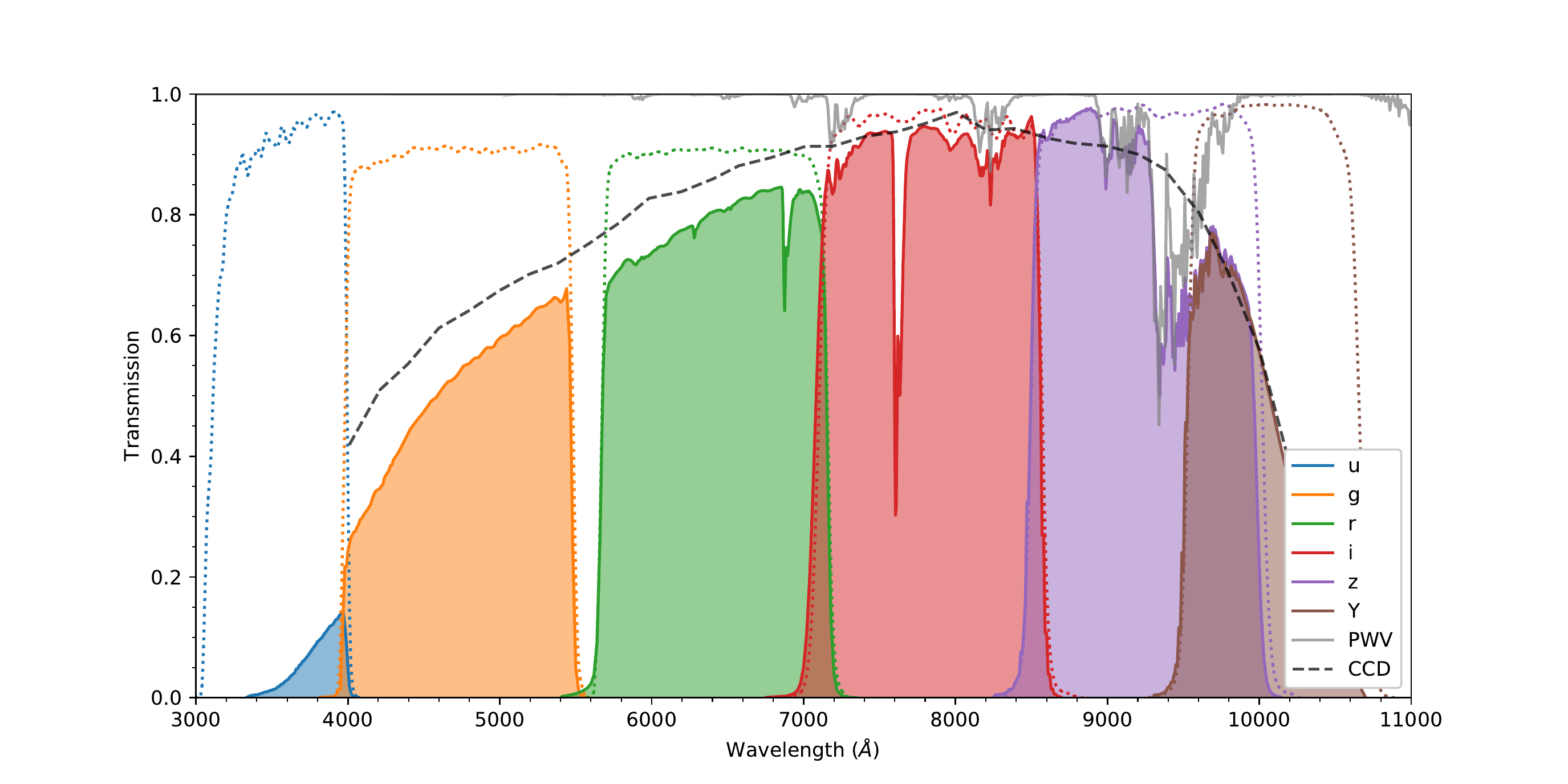}
	\caption{The optical throughput of DECam in the $ugrizY$ band-passes.
    The MzLS z-band filter is the same design as the DECam filter.
    Absorption due to PWV (solid gray) is for a fiducial atmosphere over KPNO (PWV $\approx 4$ mm) in addition to the QE response of the CCD (dashed black).
        The transmission from these components is multiplied with the transmission of each optical filter (dotted) to determine the transmission of each individual band-pass (shaded).
    The filter transmission and detector QE are smooth, but PWV absorption completely determines the sensitivity in the red half of the z-band filter.
    The $i$-band transmission is also affected by PWV absorption, but the effect is a significantly smaller fraction of the transmission.
    }
    \label{fig:mzls_filters}
\end{figure*}

The Mayall $z$-band Legacy Survey (MzLS) is one of the three public surveys that comprised the DESI Legacy Imaging Surveys\footnote{\url{http://legacysurvey.org}} \citep{Dey19}.
Using the Mosaic-3 camera on the 4-meter Mayall telescope at Kitt Peak National Observatory (KPNO) \citep{Dey14, Dey16}, MzLS observed approximately $5,100$ deg$^2$ of the sky in the $z$-band.
Mosaic-3 represents a significant upgrade from the Mosaic-1 and Mosaic-2 cameras \citep{Dey14}, with a 500~$\mu$m thick CCD that significantly increased the quantum efficiency out to 1~$\mu$m.
The $z$-band Mosaic-3 filter, shown in Figure \ref{fig:mzls_filters}, was chosen to be very similar to the $z$-band filter of the Dark Energy Camera \citep[DECam][]{Flaugher15} on the CTIO 4~m Blanco telescope to allow for standard comparisons across hemispheres.
After a commissioning phase for Mosaic-3 from 2015 October--December, the MzLS survey formally ran from 2016 February 2 through 2018 February 12.
There are 60,431 exposures recorded in the MzLS archive, with 60,403 exposures with a non-zero recorded zero point.

In order to increase the observable depth, MzLS used a 3-pass strategy to tile the sky \citep{Burleigh20}.
To ensure the accurate calibration of each exposure, the first pass of observations were performed exclusively under two conditions; clear skies with a transparency $\geq 90\%$, and a seeing better than $1.3$\arcsec.
The second pass was performed when at least one of these conditions were met, and the third pass was performed with no requirements.
This approach guaranteed at least one high-quality observation at each location in the sky that can be used to calibrate the photometry across the entire survey footprint.

MzLS observations taken during the second and third passes were calibrated by directly matching to overlapping observations taken during the first pass.
The zero points were then determined individually for each CCD by measuring the instrumental magnitude of each source and matching the source to photometry from the first Pan-STARRS data release \citep[PS1 DR1;][]{Schlafly12}.
To facilitate this comparison, a subset of PS1 DR1 sources were selected as calibrators and their colors were compared empirically to create a map between the PS1 and MzLS instrumental systems \citep[][Eq. 6]{Dey19}.
Thus while the MzLS natural system itself is absolutely calibrated to PS1,
the individual image-by-image zero points are in the MzLS natural system.

\subsection{PWV} \label{sec:gps_pwv}

\begin{figure*}
	\plottwo{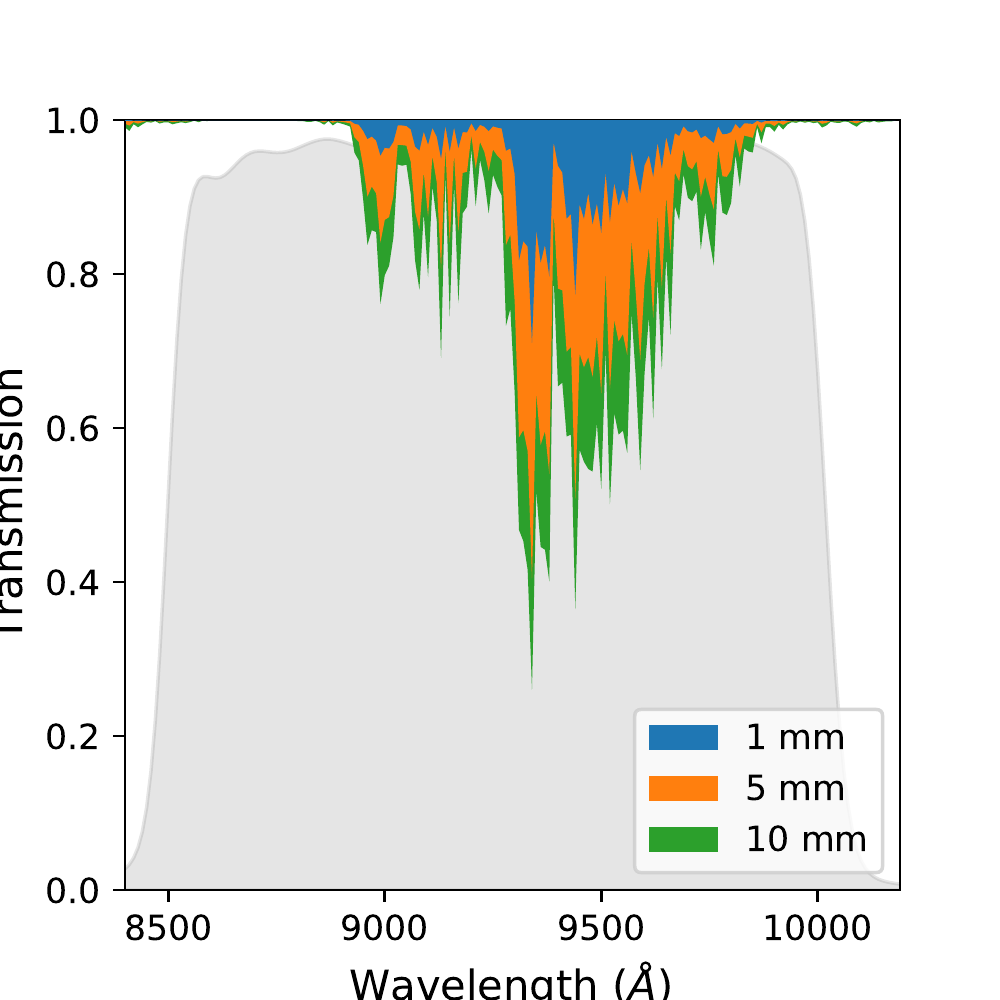}{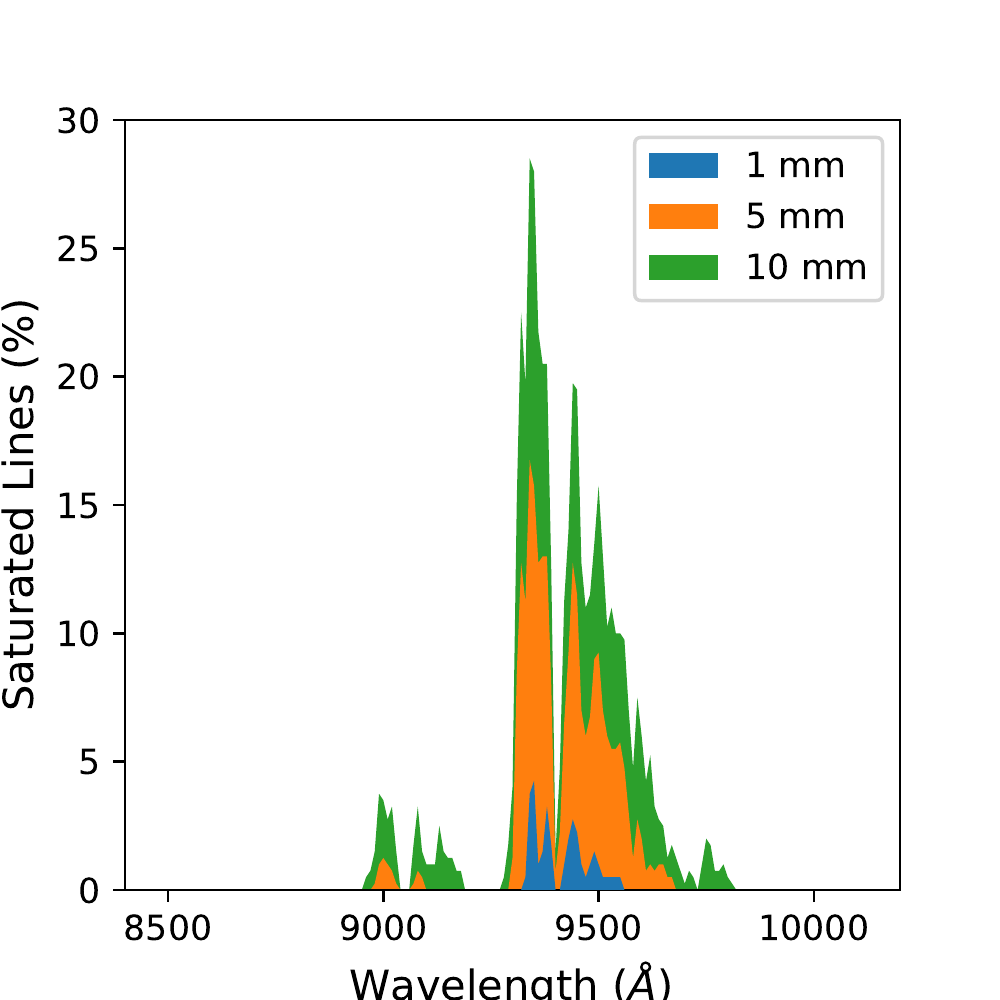}
	\caption{(left) the modeled atmospheric transmission due to PWV for 1~mm (blue), 5~mm (orange), and 10~mm (green) of PWV along line of sight. (gray) the MzLS $z$-band filter response.  As the PWV concentration increases, PWV absorption lines begin to saturate. (right) percentage of saturated lines as a function of wavelength using a 0.05~nm resolution atmospheric model for each PWV concentration.}
	\label{fig:pwv_variability}
\end{figure*}

In 2015 March, we installed a dual-band GPS receiver on the WIYN 3.5~meter telescope building at KPNO to monitor the local PWV column density \citep{Perrefort19}. Estimation of the PWV concentration from the resulting meteorological data is provided by the SuomiNet project \citep{Ware00}. With the addition of pressure, temperature, and relative humidity measurements, SuomiNet estimates the PWV column density at zenith by measuring the delay of dual-band GPS signals traveling through the atmosphere. For KPNO, these values are compiled by SuomiNet at thirty-minute intervals.

The application of GPS receivers to monitor PWV works by measuring the relative phase shift between two GPS frequencies emitted through the atmosphere \citep[for details on the methods, see][]{Bevis94, Bevis92}.
The magnitude of the delay imposed on each signal is directly related to the frequency-dependent index of refraction along the optical path.
This delay, known as the Zenith Total Delay, can be separated into a wet (PWV driven) and dry (non-PWV) component through the use of atmospheric models \citep{Tralli90}.
Although this approach is a recent advent in astronomy \citep{Braun01, Dumont01, Nahmias04, Blake11}, the use of GPS to measure PWV has a longer history in meteorology \citep{Bevis92} and has undergone continual improvement with a focus on minimizing the estimated uncertainty, often achieving levels $\leq 2$~mm \citep{Moore15, Shangguan15, Sapucci19}.

\begin{figure}
\plotone{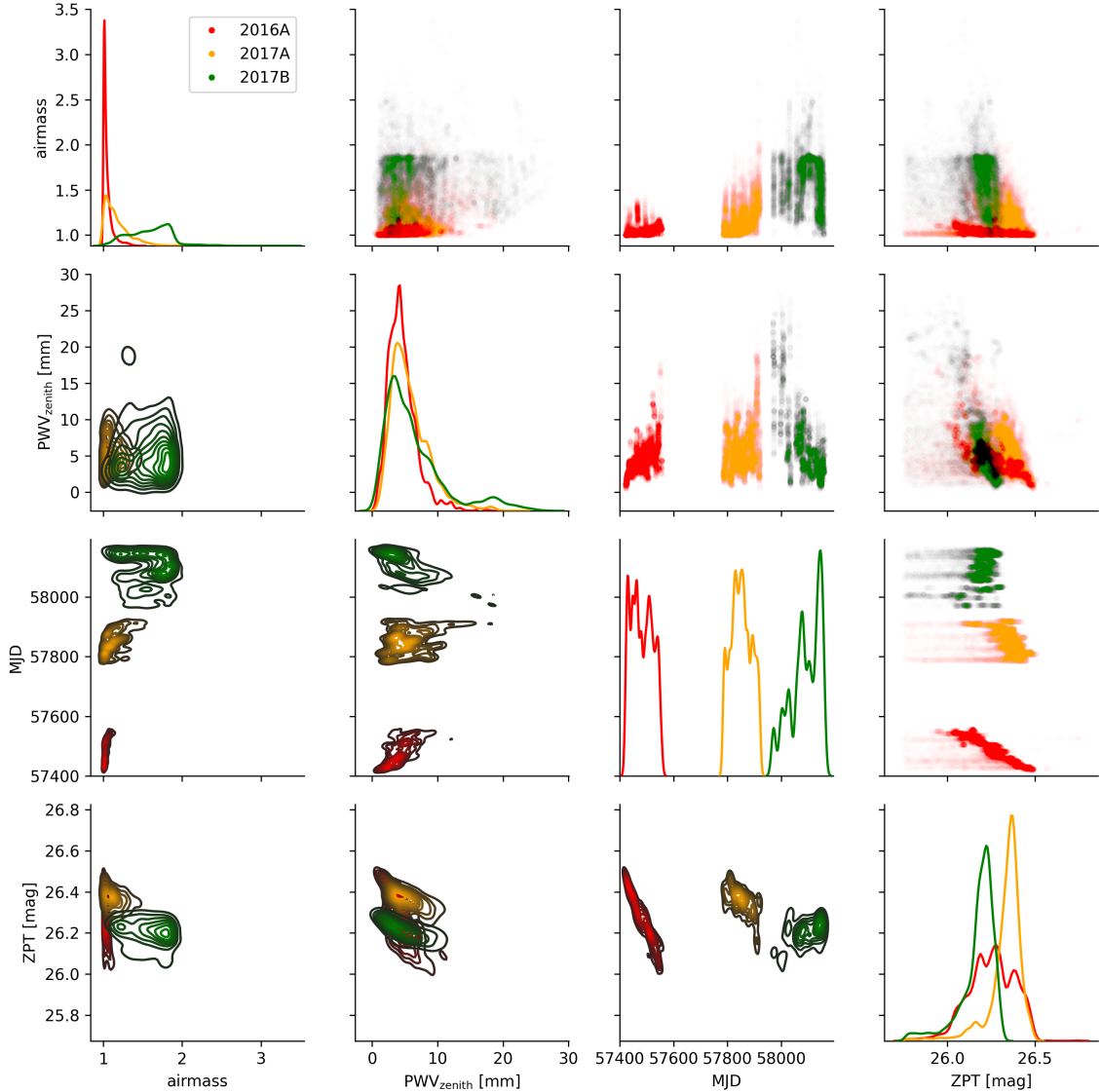}
\caption{The airmass, \pwvzenith, MJD, and zero point distribution for the MzLS data.
The points and lines are color-coded by season: (red) 2016A, (yellow) 2017A, (green) 2017B.
The upper-right are scatter plots, while the lower-left are contours from gaussian kernel density estimates.
The diagonal plots are normalized gaussian kernel density estimates of that intersecting parameter.
Note that the airmass distribution in the last semester of the survey was much wider.
The distribution of \pwvzenith was relatively the same across seasons,
but the significantly higher airmass in 2017B means that \pwvlos was systematically greater for observations in this season.
    \label{fig:airmass_pwv_mjd_zpt}
}
\end{figure}

From 2016 January through March, the barometric sensor at KPNO malfunctioned intermittently.
This caused successive non-physical spikes in PWV estimates measured during this period.
To avoid propagating these errors into this work, we ignore any PWV measurements taken over this three month period.
This removes $\sim$10,000 of the main-survey MzLS exposures.
Table~\ref{tab:mzls_data} details the number of exposures for subsets of the MzLS survey relevant for this work.
Figure~\ref{fig:airmass_pwv_mjd_zpt} shows the distribution of airmass, MJD, and zero point for the MzLS survey along with the \pwvzenith values measured from the KPNO dual-band GPS.
\pwvzenith values are comparable across the seasons, although 2017B started with a particularly high \pwvzenith during the monsoon season.
However, the airmass values in the 2017B season were significantly higher because the MzLS region extends from 100$<$RA$<$300 degrees and so observations made in the second half of the year have to reach over to access the MzLS region.
Once the full MzLS region became more visible in the last part of 2017B, the airmass of the observations came back down.

\begin{deluxetable}{lr}
\tablecaption{MzLS Exposure Counting\label{tab:mzls_data}}
\tablehead{\colhead{Set} & \colhead{Number of MzLS Exposures}}
\startdata
MzLS Survey + Commissioning & 60,431 \\
MzLS Survey & 59,552 \\
MzLS Survey in 2016A, 2017A, 2017B & 56,646 \\
MzLS Survey with KPNO PWV & 49,321 \\
$25.75 < {\rm zero\ point} < 26.75$ mag & 57,243 \\
``Clean'': $25.75 < {\rm zero\ point} < 26.75$ mag & 43,373 \\
\hspace{0.6 in} in 2016A, 2017A, 2017B & \\
\hspace{0.6 in} with KPNO PWV & \\
\enddata
\end{deluxetable}

\subsection{Telluric Transmission Models} \label{sec:telluric_models}

To manage data access for GPS measurements taken at KPNO, and to simulate the PWV transmission function, we use the \code{pwv\_kpno} Python package \citep{Perrefort19}. Using the wavelength dependent cross-section of atmospheric H$_2$O from MODTRAN \citep{modtran}, \code{pwv\_kpno} scales the atmospheric transmission function according to the effective PWV concentration along line of sight. The model provides simulations exclusively for PWV absorption, and does not include effects introduced by other atmospheric components. In Figure \ref{fig:pwv_variability} in the left panel, the water vapor transmission model for 1, 5, and 10~mm values of PWV are plotted over the MzLS $z$-band. In the right panel of Figure \ref{fig:pwv_variability} we show the percent of saturated lines as a function of wavelength.

To model Rayleigh scattering and oxygen absorption, we use the TAPAS web service~\citep{Bertaux2014} to generate the atmospheric transmission function for KPNO at an airmass of 1.
This PWV absorption spectrum is effectively identical to the one using pwv\_kpno once scaled to the same PWV value.

\section{Results from MzLS} \label{sec:mzls_results}

We here analyze the dependency of the MzLS zero points and per-epoch variations in inferred magnitudes of stars as a function of their color and PWV.
Section~\ref{sec:image_zeropoints_mjd} describes the MzLS zero points and discusses how the main contribution is a gray secular trend consistent with dust accumulation on the optical surfaces.
Section~\ref{sec:image_zeropoints_pwv} presents that the majority of the remaining zero-point variation is well-explained by the GPS-measured PWV at KPNO.
Section~\ref{sec:image_zeropoints_mjd_pwv} confirms that the effects of the long-term secular MJD trend and the PWV variation are largely separable.
Section~\ref{sec:delta_z_color} examines the per-epoch observations of each star and analyzes the variation in inferred magnitude as a function of stellar color and PWV.

\subsection{A Clean Sample of Zero Points}
\label{sec:image_zeropoints_clean}

We define a ``clean'' sample of zero points as being only from the 2016A, 2017A, and 2017B semesters.
We further restrict the clean sample to have zero points between 25.75--26.75~mag.
The minimum cut eliminates the extended tails of lower values of zero points from images taken in higher-opacity conditions, while the maximum cut eliminates a few outliers at greater than 26.75~mag that are clearly anomalous.

The clean sample shows a normalized median absolute deviation (NMAD)\footnote{NMAD is normalized such that a Guassian distribution with a $\sigma=1$ will have a NMAD$=1$.} of 131~mmag.

\subsection{Image Zero Points vs. MJD}
\label{sec:image_zeropoints_mjd}

\begin{figure}
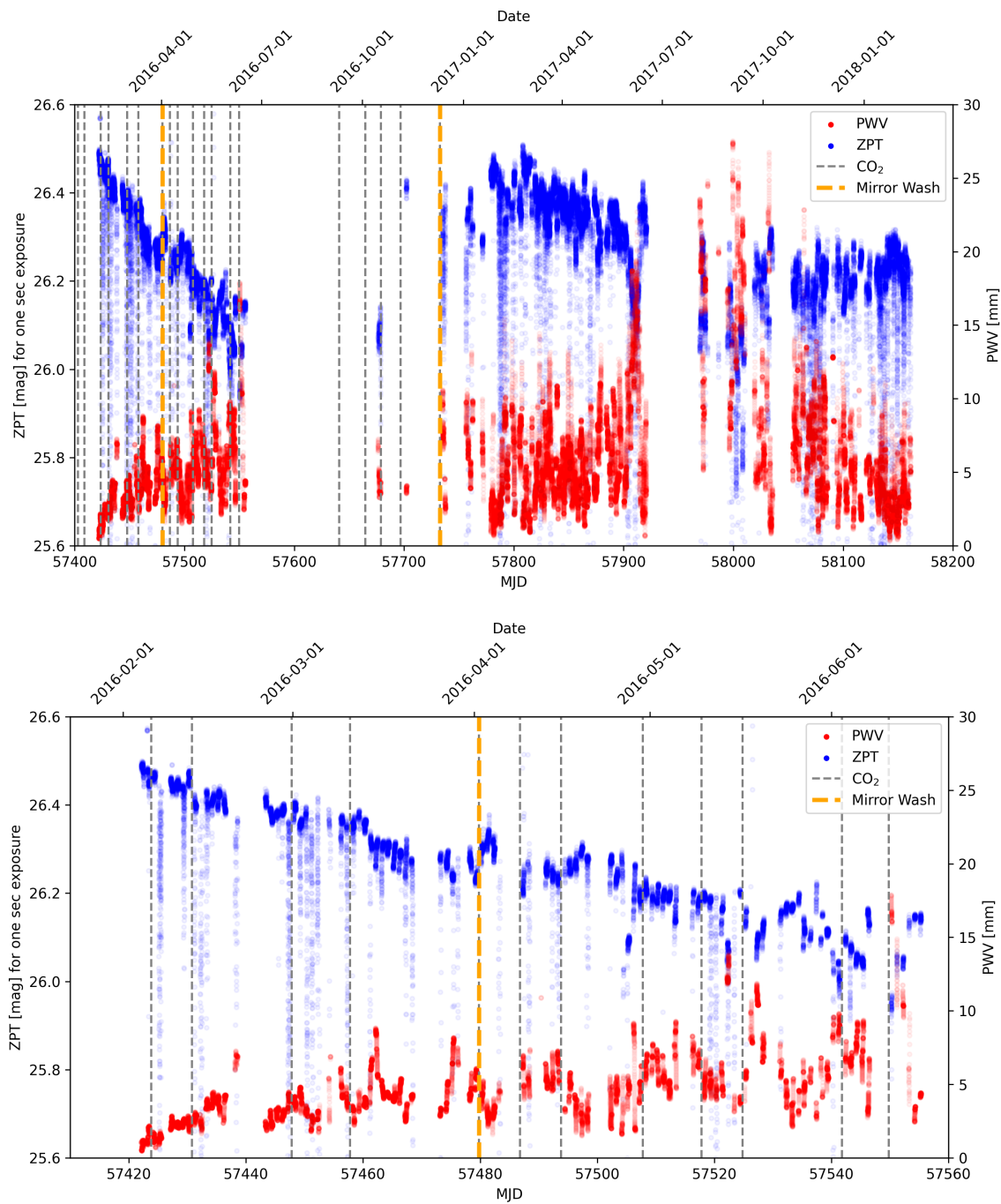

\plotone{f4.pdf}
\plotone{f5.pdf}
\caption{
(top) The MzLS image zero points (blue) show a clear secular trend with time.  There is also a clear anti-correlation between the zero point and the GPS-measured amount of PWV (red), but that accounts for less of the variation than the secular time evolution.  However, note the very clear decreases in sensitivity associated with the very high PWV levels at MJD$\sim$57910 and 57980--58020.
The cleanings of the Mayall optics (CO$_2$, gray dashed) and mirror washes (gold, dash-dot) compared with the zero point and PWV measurements.
There is no clear association with the CO$_2$ washes.
(bottom) Expanded view of the first season (2016A).  The full mirror wash potentially lead to an improvement in sensitivity on MJD=57480, but not nearly significant enough to return the zero point at the beginning of the season.}
\label{fig:zpt_mjd}
\end{figure}

The most immediately apparent correlation of the MzLS image zero points is a secular trend with MJD. Figure~\ref{fig:zpt_mjd} shows the distribution of image zero point to MJD. This trend appears to reset on an annual basis.
We first speculated that this was due to an accumulation of contaminants (``dust'') on the Mayall 4-m optical surfaces.
\citet[][Fig. 9]{Burke18} find a similar linear decrease in sensitivity for the CTIO Blanco 4-m and DECam during the Dark Energy Survey; periodic washings restored the sensitivity.
However, when we compared to the Mayall CO$_2$ cleaning schedule, Fig~\ref{fig:zpt_mjd}, we did not find a clear correlation with the cleaning schedule.  There was a potential improvement in sensitivity when the main mirror was hand washed, but the change was much smaller than the decrease over a season.

We do not have an explanation for why the slope of the decrease in ZPT with time is steeper in 2016 than in 2017.
We speculate that different amounts of dust in the air or differences in air flow across optical elements may explain the difference.
However, the effect is clear, on a longer time scale than PWV variations, and we can model it.

We estimate a simple linear model for a decrease in sensitivity with time.  The model assumes that sensitivity is reset at the beginning of 2016A and 2017A:
\begin{equation}
{\rm zp} = {\rm zp}_{\rm ideal} - \alpha~{\rm phase}
\label{eq:zpt_mjd}
\end{equation}
with ${\rm zp}_{\rm ideal} = (26.50, 26.51)$~mag and $\alpha=(2.320, 0.524)$~mmag/day.
The ${\rm phase}$ is defined with respect the beginning of the observing years: (2016A: 57419, 2017A+2017B: 57785)
Accounting for this linear secular evolution results in an NMAD of corrected zero points of 58.6~mmag.

\begin{figure}
    \plotone{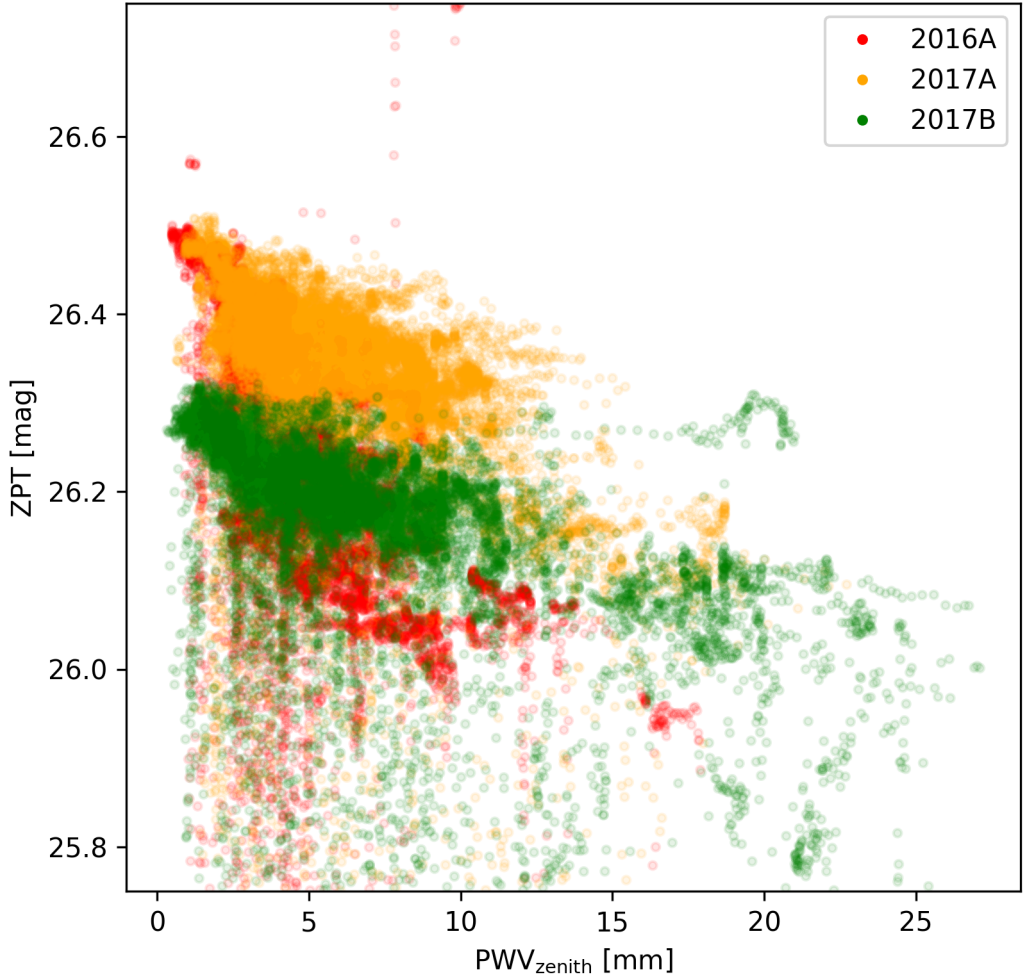}
    \caption{(left) The zero-point sensitivity for the $\sim$43,373 ``clean'' $z$-band exposures in the MzLS survey.  The color coding gives the MJD: briefly, red is 2016A, green is 2017A, yellow is 2017B.  These zero points are normalized to a one-second exposure and are {\it not} corrected for the MJD secular trend visible in Figure~\ref{fig:zpt_mjd}.}
    \label{fig:zpt_pwv_clean_mjd}
\end{figure}

\begin{figure}
\plotone{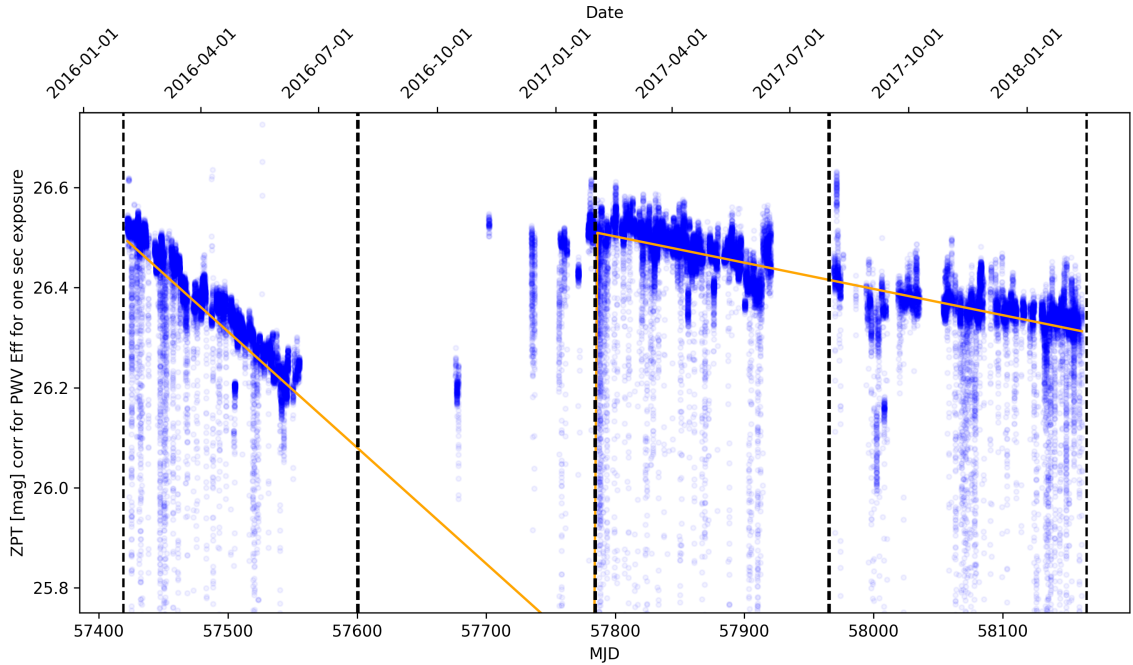}
\caption{(blue) PWV-corrected MzLS zero point vs. MJD.  If we remove the estimated effect of PWV, we get an even cleaner signal of the secular evolution of MzLS sensitivity with MJD.  Note, in particular that the continuation of the trend in 2017A is now clearer in 2017B.  In Figure~\ref{fig:zpt_mjd} the 2017B semester appeared to flatten out from the simple linear trend of 2017A.  But 2017B was a drier semester than 2017A and thus there was less absorption due to PWV and the zero points are higher.  Once the difference in PWV is taken into account, the linear dependence on ZPT with MJD is more clear.
(orange) Linear time evolution models per year: -2.320 mmag/day in 2016: -0.524 mmag/day in 2017.
We attribute this secular time evolution to an accumulation of contaminants on optical surfaces in the telescope.
Note that during the highest PWV events, e.g., MJD$\sim$58010, the zero point loss of sensitivity is not fully explained by PWV.  We hypothesize that high PWV is also associated with increased gray absorption from condensed water vapor.
}
\label{fig:zpt_mjd_model}
\end{figure}

\subsection{Image Zero Points vs. PWV}
\label{sec:image_zeropoints_pwv}

\begin{figure}
\plotone{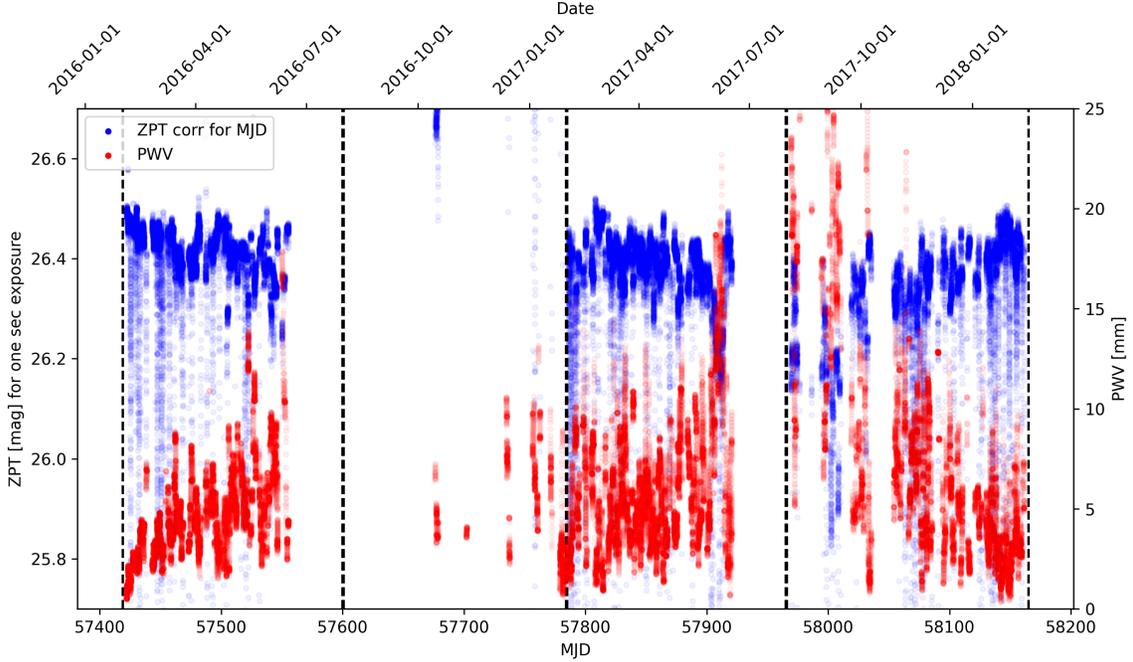}
\caption{(blue) The MzLS zero point values corrected for secular MJD trend, but {\it uncorrected} for PWV.  (red) The reported KPNO PWV measurements.  The significant dips in MJD-corrected zero points are clearly anticorrelated with significantly elevated column densities of PWV.}
\label{fig:zpt_pwv_vs_corr_mjd}
\end{figure}

After correcting for the secular component of zero-point variation, the remaining variation in the zero points is well-explained by the variation in PWV (Figure~\ref{fig:zpt_pwv_vs_corr_mjd}).
The GPS-based PWV measurement is for the PWV at zenith, ${\rm PWV}_{\rm zenith}$, as the combination of measurements of delays from several satellites spread across the sky with local meteorological data (pressure, temperature, humidity).
Any given observation with the Mayall 4-m at an angle away from zenith will look through more water vapor.  If we assume that the PWV is distributed as a uniform slab in the atmosphere, then the amount of water vapor along a given line of sight, \pwvlos, will scale with the airmass, $X=\sec({\rm zenith}~{\rm angle})$,
\begin{equation}
    \pwvlos = \pwvzenith X
    \label{eq:pwv_los}
\end{equation}

However, the absorption does not scale linearly with the amount of water vapor.
Water vapor absorption is a rich mix of lines, many of them at relatively high optical depth even at just a few mm of PWV.
Thus the water vapor absorption displays a non-linear relationship between total absorption in a given wavelength window and PWV column depth.
Once an individual line saturates, increasing water vapor column depth no longer leads to additional absorption, so absorption increases with increasing PWV more slowly than linearly.
\citet{Wade88} estimated that one could account for this saturation by calculating an effective line-of-sight PWV by scaling by airmass to the $\delta=0.6$ power:
\begin{equation}
	\pwvwade = \pwvzenith {X}^{0.6}
\end{equation}
This correction allows for a single linear coefficient to describe the increase in absorption (decrease in zero point, zp) with airmass:
\begin{equation}
	{\rm zp} = {\rm zp}_{\rm ideal} - \beta~\pwvzenith X^\delta
\label{eq:zp_pwv_wade}
\end{equation}
We find that for an $\delta = 0.6$, $\beta = 12.1$~mmag/mm provides an improved fit for translating the measured zero points to a idealized constant zero point.

While \citet{Wade88} gave the scaling as just in airmass X$^{0.6}$, physically the saturation effect should follow total water vapor along the line of sight, \pwvlos;
i.e., looking through 10~mm of water vapor should yield the same absorption profile whether that's at \pwv=10~mm at an airmass of 1 or \pwv=5~mm at an airmass of 2.
We set the normalization PWV$_{\rm norm}$ = 2~mm at the lowest value of \pwv regularly measured in the data and model the zero point as
\begin{equation}
	{\rm zp} = {\rm zp}_{\rm ideal} - \beta~\left( \frac{\pwvzenith}{{\rm PWV}_{\rm norm}} X \right)^\delta
\label{eq:zp_pwv_eff}
\end{equation}
For this definition of effective PWV, we find $\delta = 0.6$ makes the zero-point dependence largely linear with $\beta = 31.9$~mmag/mm$^{0.6}$.
This corrected effective PWV prescription corrects the zero point variation better than the Wade prescription (see Table~\ref{tab:coefficients_nmad}).

Using detailed water absorption and stellar spectral models, we can directly calculate the expected absorption of a given spectral energy distribution (SED) due to a given amount of water vapor along the line of sight.
Any observation contains stars of many different stellar types and the calibration will represent an aggregation across the range of stars.
The full details of the calibration process depend on choice of signal-to-noise weighting and the brightness-color distribution of objects, but we here illustrate the effect by using a single representative stellar type.
For the MzLS data we find that K9 star can represent the typical calibration and variation.
If we are calibrating the zero point using a K9-type star, then
\begin{equation}
    {\rm zp} = {\rm zp}_{\rm ideal} - \gamma\ {\rm K9}_{\rm absorption}\left(\pwvlos\right)
\label{eq:zp_pwv_k9}
\end{equation}
where ${\rm K9}_{\rm absorption}(\pwvlos)$ is the predicted absorption of a K9 spectrum
through \pwvlos as integrated over the relevant passband for which ${\rm zp}$ is being determined.
If we are correctly modeling all of the relevant physics, $\gamma$ should be equal to 1.

\begin{figure}
    \plottwo{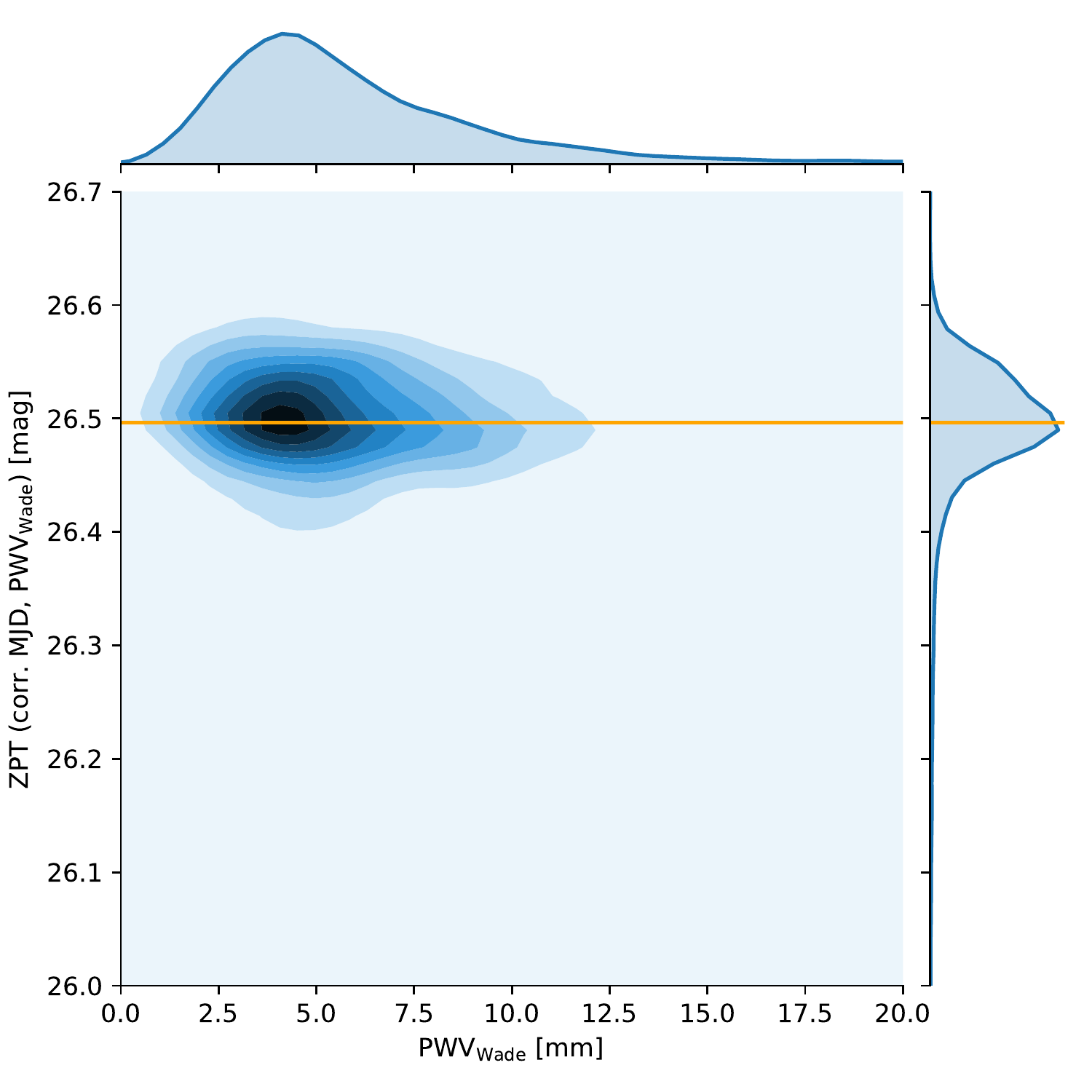}{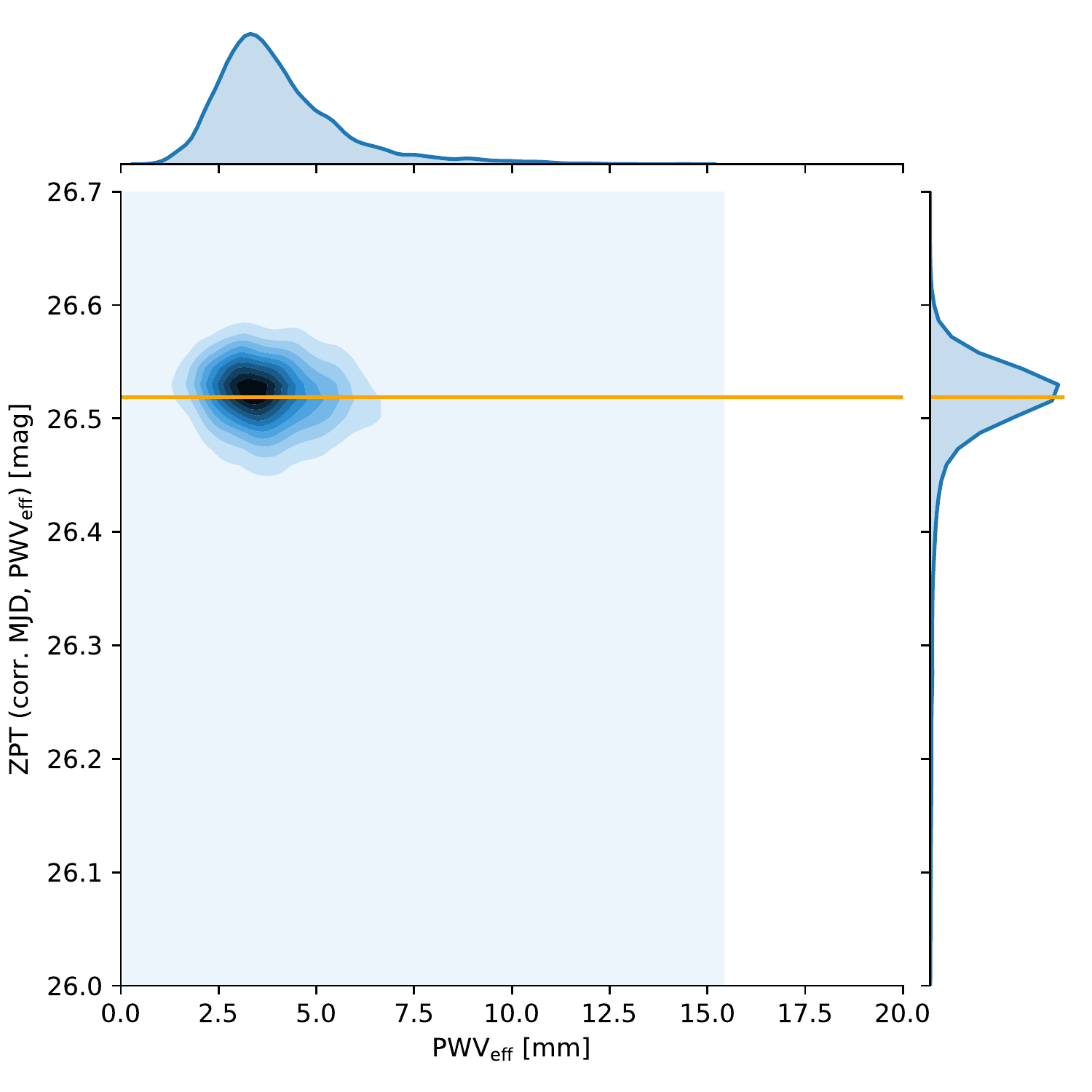}
    \plottwo{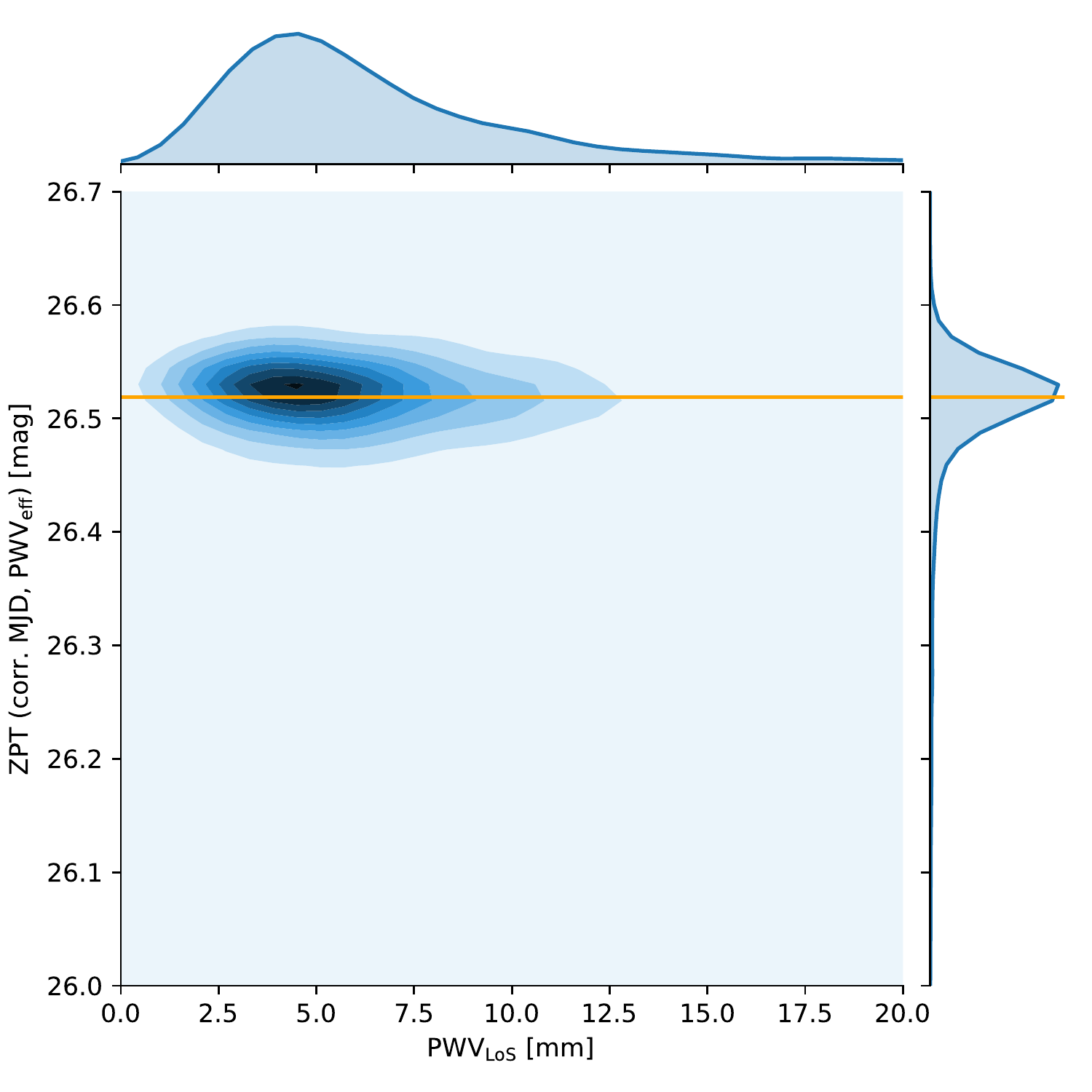}{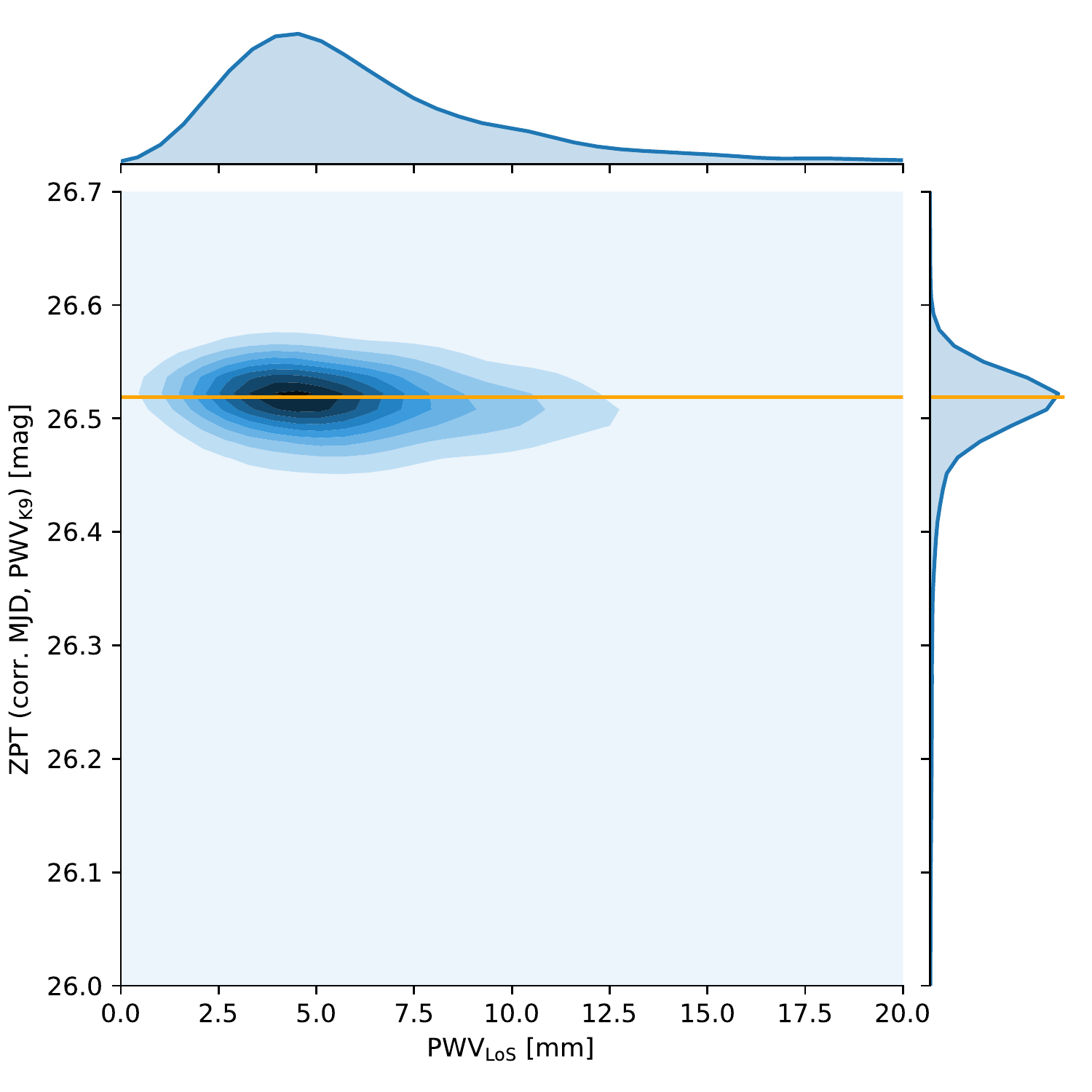}
        \caption{MzLS MJD-corrected and PWV-corrected zero point for four different methods of calculating effective PWV and correcting for it.
    (top left) corrected by $0.0121 \times \pwvzenith (X)^{0.6}$; (top right) $0.0318 \times (\pwvzenith X)^{0.6}$;
    (bottom left) same but plotted against \pwvlos; (bottom right) corrected by the absorption expected for a K9 stellar model.
    The orange horizontal lines are the median of each corrected zero point distribution.
    }
    \label{fig:zpt_pwv_mjd}
\end{figure}

If we use this prediction for a K9 star to model the zero point variation for the MzLS data and fit for $\gamma$ in Equation~\ref{eq:zp_pwv_k9}, we find $\gamma = 1.097 \pm 0.008$.
While this linear fit value is formally well-constrained, a value of $\gamma=1$ yields very similar NMAD of the residuals (29.6~mmag for $\gamma=1.097$ and 30.6~mmag for $\gamma=1$).
Because using a $\gamma$ fit coefficient is not particularly well-motivated, we remove this degree of freedom (by fixing $\gamma=1$) for the stellar+PWV model corrections used in the rest of this paper.
Figure~\ref{fig:zpt_pwv_mjd} compares of the distribution of the residuals after fitting with each of Equations~\ref{eq:zp_pwv_wade}, \ref{eq:zp_pwv_eff}, and \ref{eq:zp_pwv_k9} (with $\gamma=1$).

\subsection{Image Zero Points vs. (MJD, PWV)}
\label{sec:image_zeropoints_mjd_pwv}

The previous sections presented an iterative fit with ${\rm zpt}_{\rm mjd}$ first calculated based on the zero point corrected by the fit dependency on MJD, and then fit for $\gamma$.
We here jointly fit for a per-season MJD dependence and a PWV dependence.  Due to the paucity of the data in 2016B, we ignore data from that semester.

We find that the zero points are well explained by the K9 model, which is a model that is directly physically motivated by stellar spectra and PWV absorption profiles:
\begin{equation}
    {\rm zp} = {\rm zp}_{\rm ideal} - \alpha~{\rm phase} - \gamma~{\rm K9}_{\rm absorption} \left( \pwvlos \right)
\end{equation}
with ${\rm zp}_{\rm ideal} = (25.3421, 26.4312, 26.3874)$~mag, $\alpha=(2.087, 0.294, 0.406)$~mmag/day, and $\gamma=(0.9963, 1.0863, 1.0968)$.
The triplets of quantities refer to the seasons (2016A, 2017A, 2017B).
The ${\rm phase}$ is defined with respect the beginning of the observing years: (2016A: MJD 57419, 2017A+2017B: MJD 57785).
Note that the $\alpha$ from the \pwvlos and \pwveff models are consistent.

The zero point variations can also be explained by the simple \pwveff model:

\begin{equation}
    {\rm zp} = {\rm zp}_{\rm ideal} - \alpha~{\rm phase} - \beta~\pwveff
\end{equation}
with ${\rm zp}_{\rm ideal} = (25.3450, 26.4330, 26.3812)$~mag, $\alpha=(2.088, 0.281, 0.415)$~mmag/day, and $\beta=(40.80, 43.12, 40.87)$~mmag/mm$^{0.6}$.

The \pwvwade is also successful, but not as good as our \pwveff definition.
The difference is particularly noticeable in the 2017B season of MzLS. In this season the median airmass was significantly higher because the MzLS fields are not overhead at the beginning of the B semesters.
But the the median PWV (after the first few weeks) was significantly lower.
Thus the \pwvlos distribution was similar to other semester but the \pwvwade distribution was higher.
Thus \pwvwade the nominal zero point correction coefficient fit on a per-season basis was 14\% lower between 2017A and 2017B to account for this increased range of airmass at a comparable \pwvlos, while the zero point correct coefficient was consistent for \pwveff between those two semesters.

\begin{figure}
    \plotone{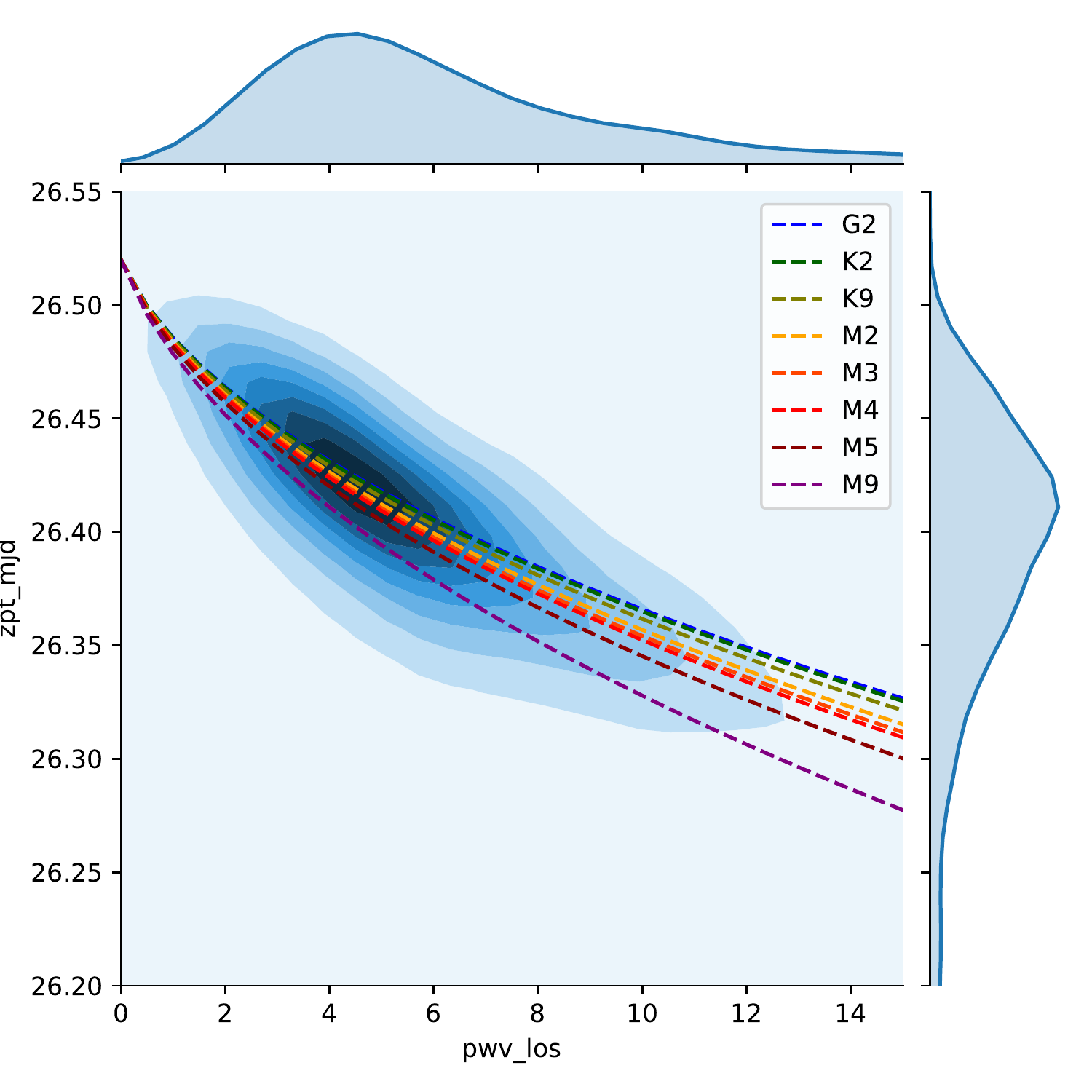}
	\caption{
    MzLS MJD-corrected zero point vs. \pwvlos.
    Overlaid in dashed lines are the expected change in $z$-band transmission when observing different stellar types:
    G2 (blue), K2 (dark green), K9 (olive), M2 (orange), M3 (orange-red), M4 (red), M5 (dark red), and M9 (purple).
    }
	\label{fig:zpt_pwv_model}
\end{figure}

Figure \ref{fig:zpt_pwv_model} shows the expected zero-point offsets as a function of PWV for a different stellar types overlaid with the MzLS zero points.
The calculation of these offsets is described in detail in Section~\ref{sec:stars}.
Table~\ref{tab:coefficients_nmad} summarizes the coefficients for the different models for zero point based on MJD and PWV.

\begin{deluxetable}{lrrrrrr}
\tablehead{
\colhead{Type} & Period & \colhead{${\rm zp}_{\rm ideal}$ [mag]} & \colhead{$\alpha$ [mmag/day]} & \colhead{$\beta$ [mmag/mm$^{0.4}$]} & \colhead{$\gamma$} & NMAD [mmag] \\
}
\tablecaption{Zero point, MJD, PWV coefficients, NMAD
\label{tab:coefficients_nmad}}
\startdata
Raw        & 2016--2017   & \nodata        & \nodata        & \nodata & \nodata & 130.9 \\
MJD              & (2016, 2017) & (26.50, 26.51) & (2.320, 0.524) & \nodata & \nodata &  58.6 \\
MJD + \pwvwade   &              &                &                &    12.1 & \nodata &  42.5 \\
MJD + \pwveff    &              &                &                &    31.9 & \nodata &  29.5 \\
MJD + K9         &              &                &                & \nodata & 1.127 &  30.6 \\
Joint (MJD, \pwvwade) & 2016A & 25.289 & 2.116 & 12.63 & \nodata & 31.7 \\
                      & 2017A & 26.380 & 0.303 & 12.74 & \nodata & 33.1 \\
                      & 2017B & 26.320 & 0.426 & 10.96 & \nodata & 33.4 \\
Joint (MJD, \pwveff)  & 2016A & 25.345 & 2.088 & 40.80 & \nodata & 31.2 \\
                      & 2017A & 26.433 & 0.281 & 43.12 & \nodata & 33.1 \\
                      & 2017B & 26.381 & 0.415 & 40.87 & \nodata & 28.0 \\
Joint (MJD, K9)       & 2016A & 25.342 & 2.087 & \nodata & 0.996 & 31.2 \\
                      & 2017A & 26.431 & 0.294 & \nodata & 1.086 & 33.0 \\
                      & 2017B & 26.387 & 0.406 & \nodata & 1.097 & 27.5 \\
\enddata
\tablecomments{Ideal zero points and MJD and PWV model coefficients.  NMAD for corrected zero point using given model.  The ``MJD *'' rows fit per-year MJD terms first and then one overall coefficient for each PWV measurement.  The ``Joint *'' rows are for simultaneous fitting of coefficients for both on a per-semester basis.}
\end{deluxetable}

\subsection{MzLS Per-Exposure Magnitude Deviations Depend on Color and PWV}
\label{sec:delta_z_color}

The DESI Legacy Survey DR8 provides per-image forced-photometry.
We here explore the correlation between the per-image
measured magnitude of a star, its color, and PWV.

First, some definitions for what will be a confusing discussion of differences of differences.
If $z^i_j$ is the $z$-band magnitude for star $i$ measured on image $j$,
then we are interested in $\Delta z^i_j = z^i_j - <z^i>$, where $<z^i>$ is the
average\footnote{It's not necessarily an average.  See Legacy Survey paper for details.} magnitude based on all of the survey images.
\pwvlos$_j$ is the measured \pwvlos for image $j$.

\begin{figure}
\plotone{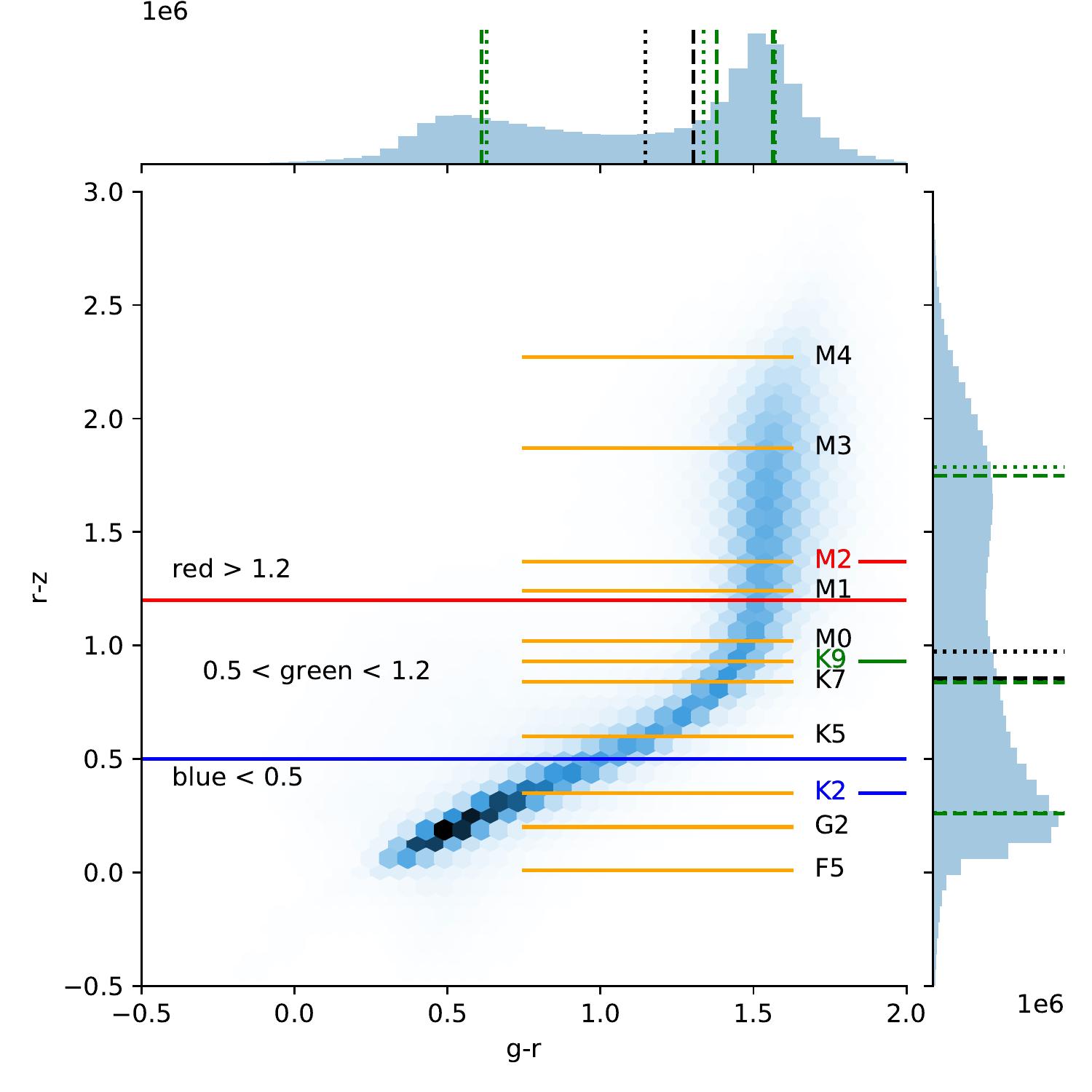}
\caption{Stellar locus for DESI Legacy Survey stars in the MzLS footprint.
No correction has been made for Milky Way extinction.
The blue, green, and red $r-z$ color regions are indicated
along with the nominal $r-z$ colors of a range of stellar types.
The top (side) panels are the projected histograms for $g-r$ ($r-z$).
The overall median (black dashed) and mean (black dotted) colors
are consistent with a K9 star.
For each color region the median and mean value are very close.
We have chosen representative stellar types of K2, K9, and M2.
Note that while the nominal mean/median red color
is notably redder than the M2 (M2.0) stellar spectrum,
it is still within the range between M2 and M3.
}
\label{fig:stellar_locus_cuts_types}
\end{figure}

The stars observed by MzLS span the range of stellar types from
F5--M4 stars (Figure~\ref{fig:stellar_locus_cuts_types}).
We here look at the relative change in $z$-band flux of specific objects under varying amounts of PWV as a function of object color.
We group stars into three broad color categories:
``blue'': $r-z \leq 0.5$~mag;
``green'': 0.5~mag $ < r-z \leq 1.2$~mag;
``red'': 1.2~mag $< r-z$.
While these color ranges are only loosely motivated, they end up being illustrative.
The ``blue'' cut is in the middle of G-type stars -- the mean color of the PS1 stars used to establish the MzLS calibration is close to a K9 (see Figure~\ref{fig:ps1_ref_cat_color}).
The ``red'' cut marks the beginning of the M-dwarf sequence,
which is where the stellar locus goes vertical in $r-z$ vs. $g-r$.
The $g$ and $r$-band magnitudes are from the Bok 90" observations
from the Beijing-Arizona Sky Survey (BASS)\footnote{\url{http://www.legacysurvey.org/bass/}} component of the DESI Legacy Survey,
and the $z$-band magnitudes are determined from the MzLS observations.

Fig.~\ref{fig:stellar_locus_cuts_types} shows a (median, mean) $r-z$ color for MzLS observed stars of ($0.85$, $0.97$)~mag.
A K9 dwarf star has a color of $r-z=0.93$~mag, which is right in between these values.
Thus even without fully recreating the calibration of the MzLS survey, one would still reasonably expect that this typical color will remain correctly calibrated through a variety of PWV conditions.
These blue$-$green and blue$-$red relations should cross through zero when the conditions match the ``average'' conditions of the set of data considered by the MzLS catalog calibration.
At higher \pwvlos, progressively bluer (redder) stars should be brighter (fainter).

The DESI Legacy Survey is divided into 0.25\arcdeg~$\times$~0.25\arcdeg ``bricks'' on the sky.
We restrict our analysis to stars in bricks that had a median of three or more exposures
contributing to the pixels in the brick: ``nexp\_z $> 3$''.
This reduced our sample from 93,610 bricks to 81,934 bricks.
We successfully retrieved photometry for 81,1156 of these bricks, which had data from 15,381 MzLS exposures.
The Mosaic-3 field of view is 36\arcmin~$\times$~36\arcmin, which means that 36 bricks fit exactly into one field of view.
The alignment is not perfect, in part by construction, and so 60 bricks should be relevant for any particular image.
We restrict our stellar sample to objects identified as stars and with $z$-band SNR $> 25$, 0.5 mag~$< g-r < 2.5$~mag, and $0.5$~mag~$< r-z < 5.0$~mag.

\begin{figure}
\plotone{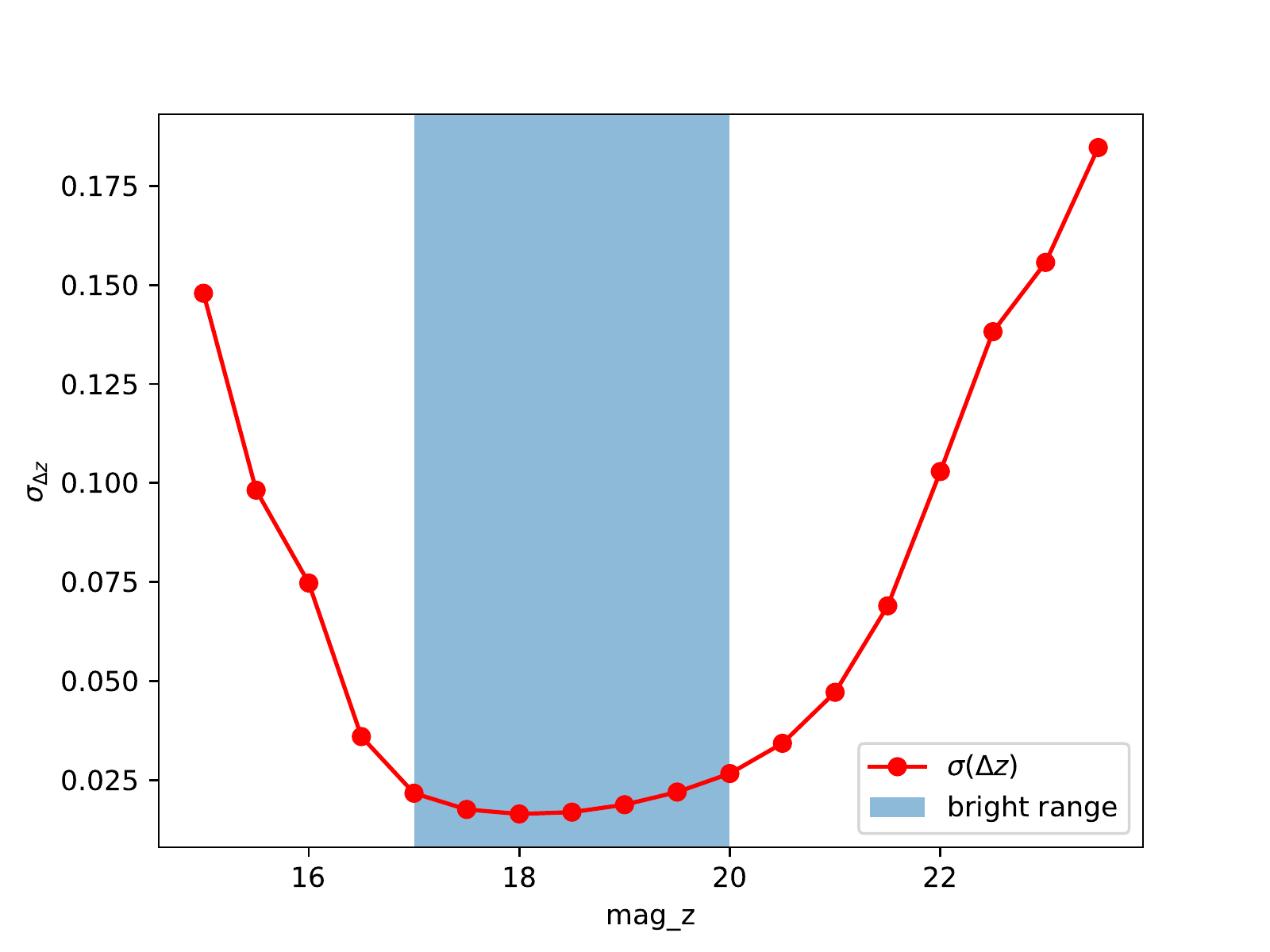}
\caption{Standard deviation of $\Delta z$ for all stars in MzLS.  (red points+line) $\sigma(\Delta z)$ binned by magnitude.  At bright magnitudes, the variance increases as measurements approach the saturated/non-linear region.  At the faint end, the variance increases as SNR decreases.  (blue rectangle) magnitude range of the ``bright'' sample we focus on in our $\Delta z$ vs. PWV analysis.}
\label{fig:delta_z_sigma_mag_z}
\end{figure}

Figure~\ref{fig:delta_z_sigma_mag_z} shows how the variance in $\Delta z$ is a function of magnitude.  Non-linearity/saturation drives up variance at the brighter end while decreasing SNR and increasing non-stellar contamination drives up variance at the faint end.  We thus further restrict our $\Delta z$ vs. PWV analysis to stars in the range of $17 < z < 20$~mag.

For each each image, and each color class, we compute the median of the distribution of $\Delta z^i_j$ values.  Each image, $j$, then has three values: $\Delta z^{\rm blue}_j$, $\Delta z^{\rm green}_j$, $\Delta z^{\rm red}_j$.

\begin{figure}
\includegraphics[width=0.3\textwidth]{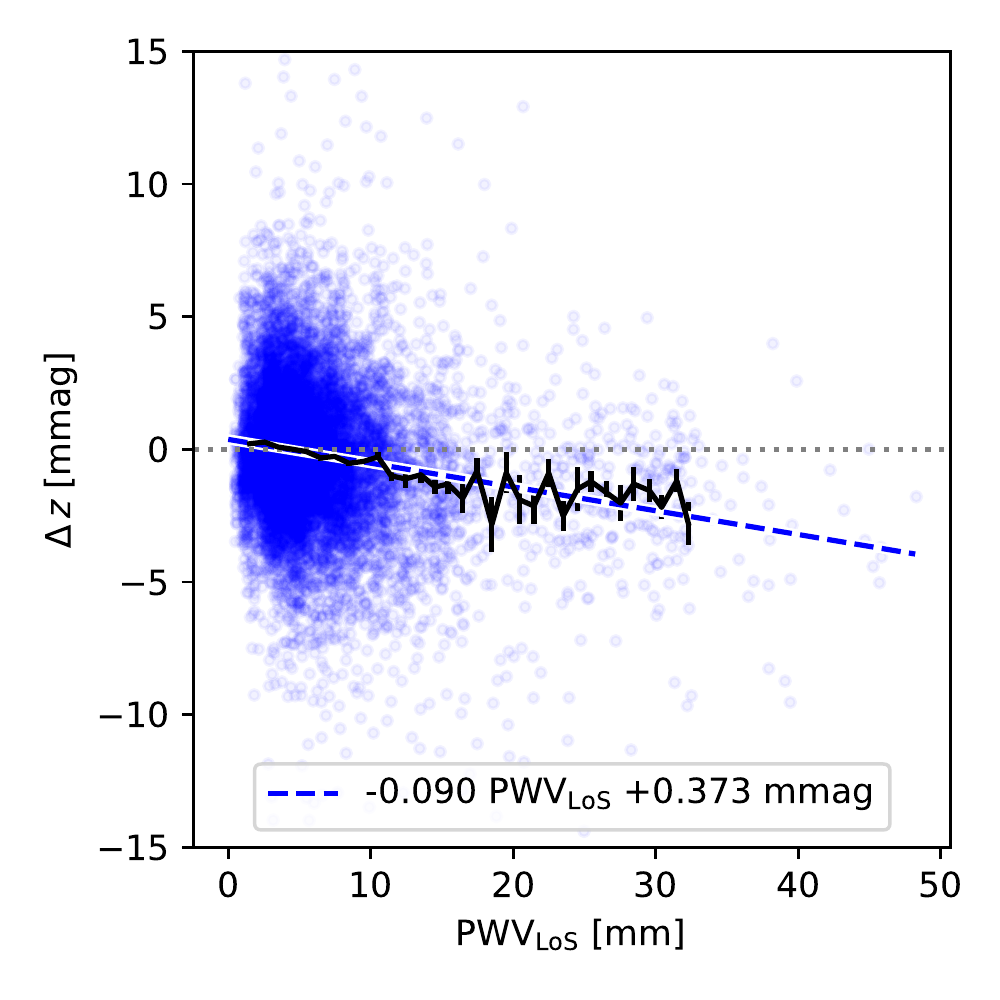}
\includegraphics[width=0.3\textwidth]{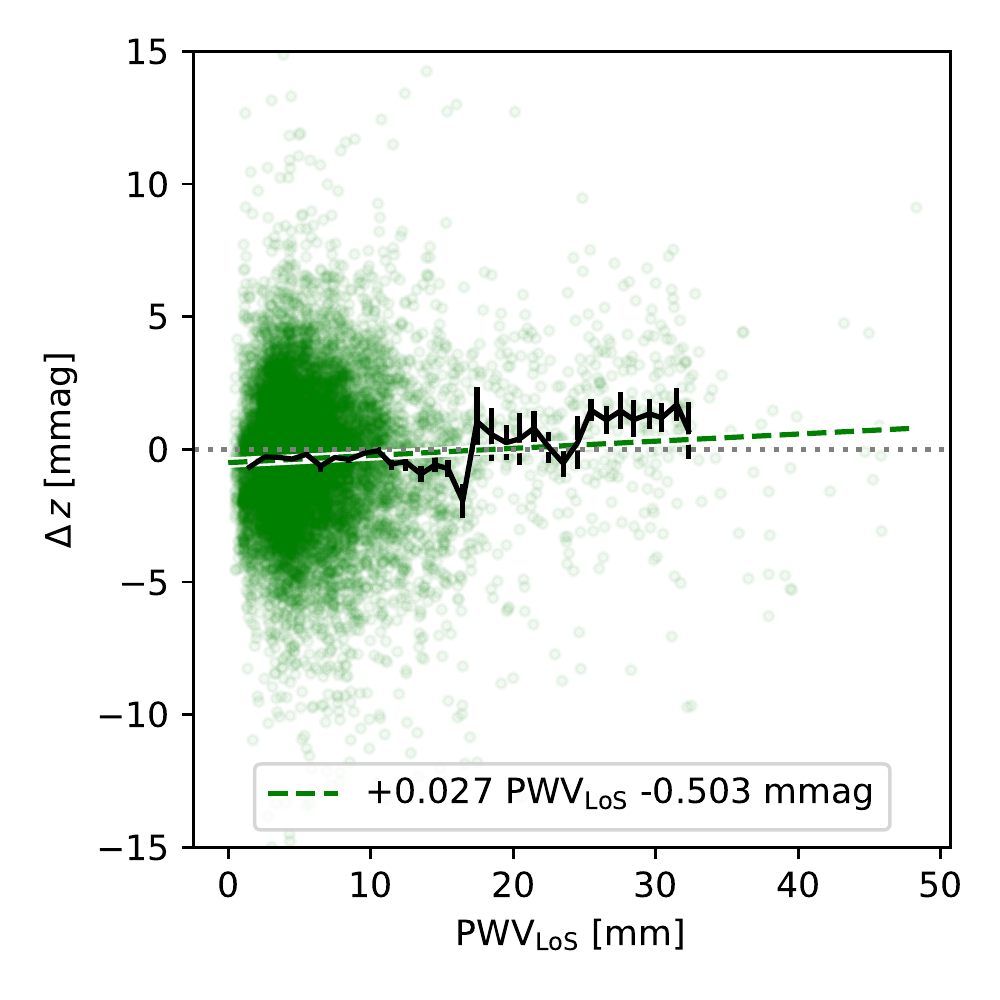}
\includegraphics[width=0.3\textwidth]{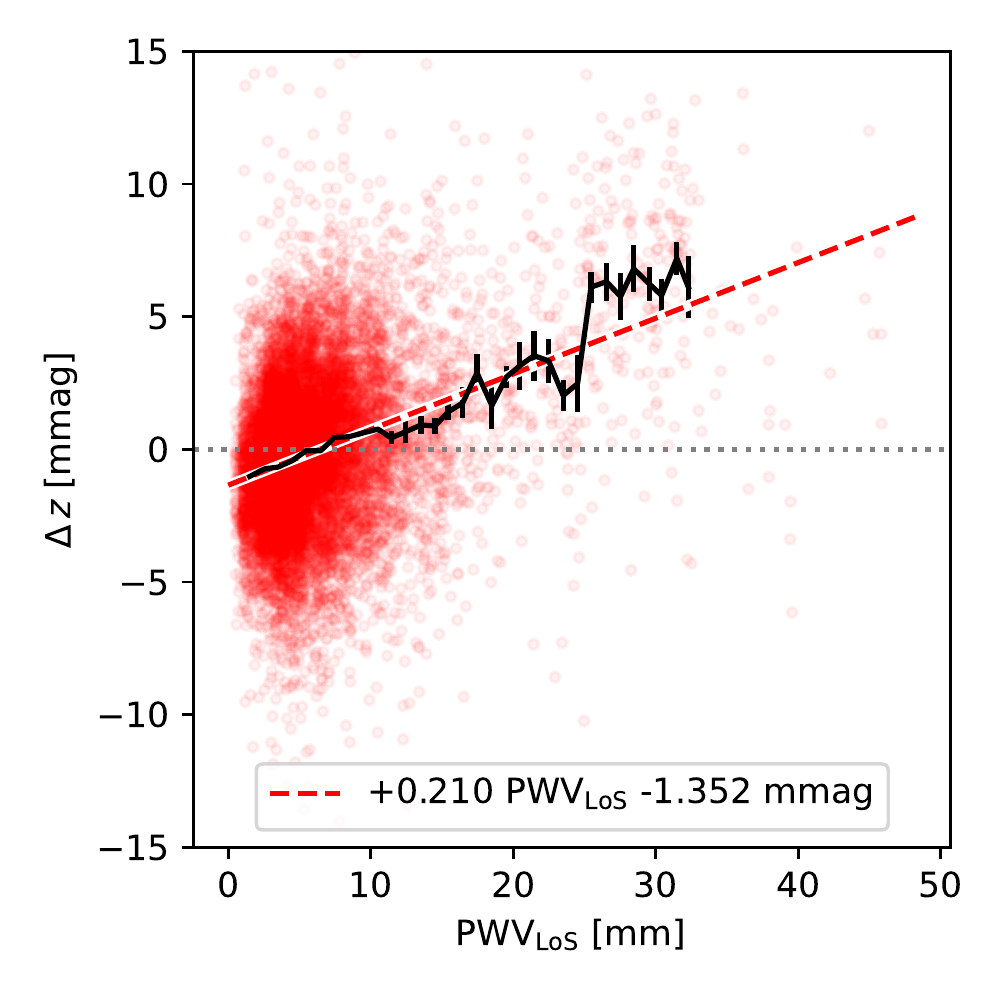}
\caption{The per-observation variation of the $z$-band magnitude of stars in three color bins versus the \pwvlos for that observation.
(left) blue: $r-z < $0.5~mag;
(center) green: $0.5 < r-z < 1.2$~mag;
(right) red: $r-z > 1.2$~mag.
Each point is the median $\Delta z$ for all of the stars in that color bin for that observation.
The median star in MzLS has a $r-z$ a bit less than 1~mag and so the ``green'' panel is largely flat.
The blue stars appear brighter at higher values of PWV,
while the red stars are clearly fainter for high values of PWV.
These binned lines intersect at the median \pwvlos of the survey: $\sim$4.83~mm.
}
\label{fig:delta_z_colors_pwv_los}
\end{figure}

Figure~\ref{fig:delta_z_colors_pwv_los} shows the trends of $\Delta z^{\rm blue, green, red}_j$ vs. \pwvlos$_j$.
As \pwvlos increases, $\Delta z^{\rm blue}_j$ becomes brighter, $\Delta z^{\rm green}_j$ is flat, and $\Delta z^{\rm red}_j$ becomes fainter.
This trend as a function of stellar type is consistent with a model
where SEDs that are bluer or redder than the average
are mis-calibrated because an effectively gray term (the zero point in the natural system) does not
properly capture the significant absorption due to PWV at the red side of the $z$ band.
If the absorption were at the blue side of the $z$ band, then the slopes would be reversed, with
blue stars being fainter at higher PWV and red stars being brighter -- relative to the green stars.
The flat trend in the green stars implies that they are representative of the effective average star used to calibrate
the MzLS observations.

They are thus flat in $\Delta z$ vs. \pwvlos.  For \pwvlos less than this value, the red stars are systematically brighter than their average.  For \pwvlos greater than this value, the blue (red) stars are systematically brighter (fainter) than their average.  The three binned lines meet at $\Delta z=0$~mag at the median \pwvlos of the survey, $\sim4.83$~mm.

\begin{figure}
\plottwo{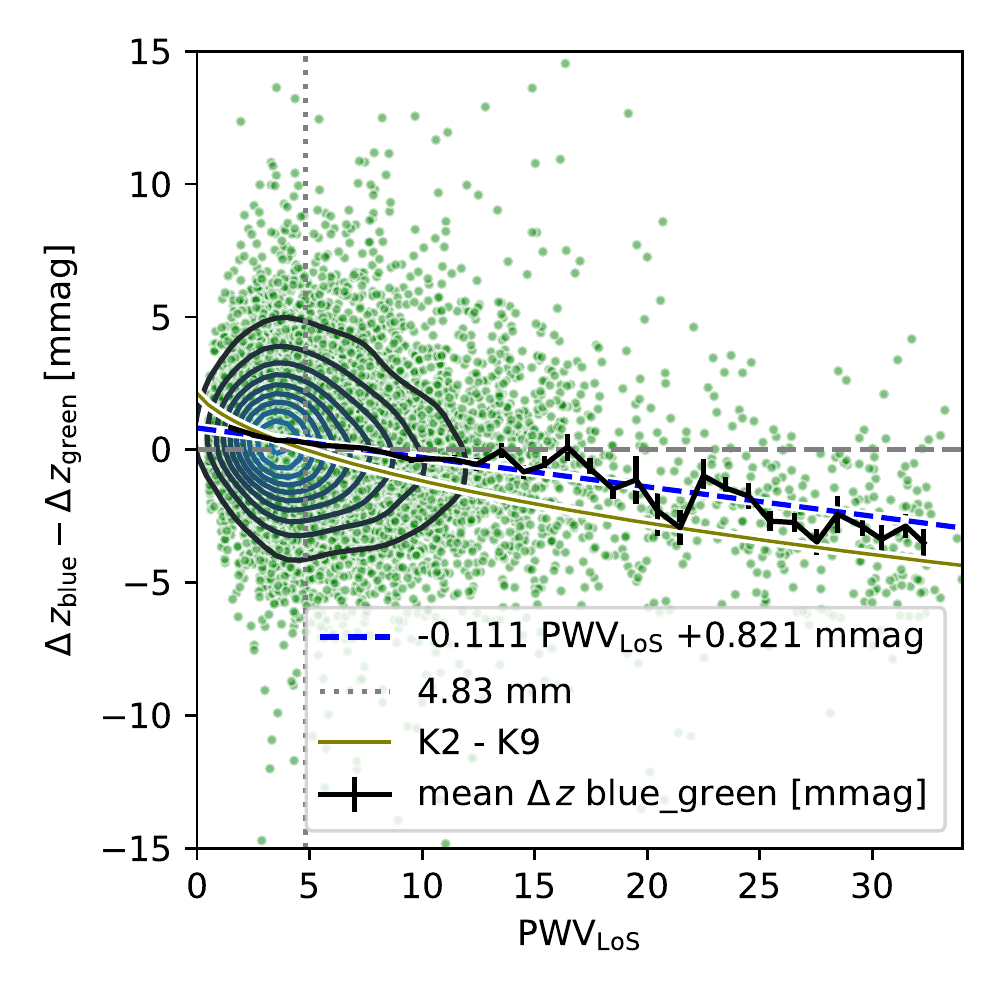}{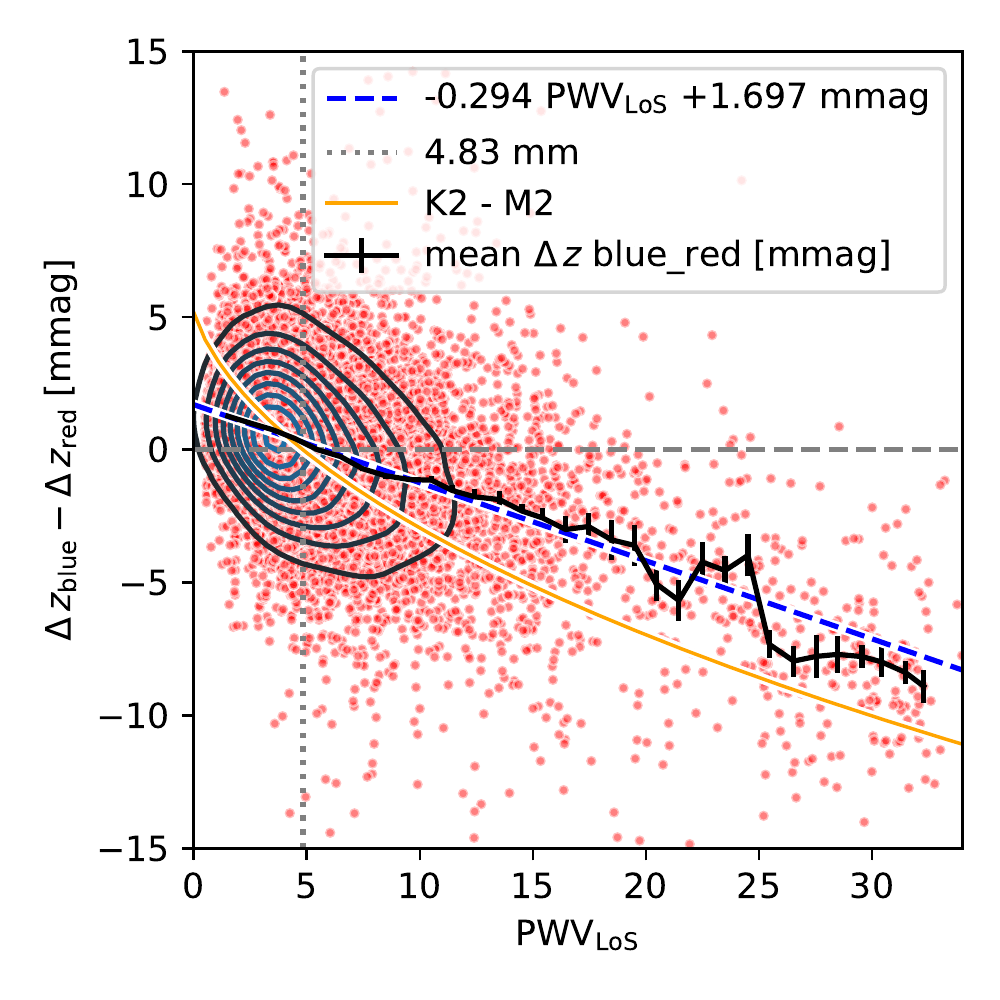}
\caption{
Difference in median $\Delta z$ between different color stars versus \pwvlos (blue points, with density estimation contours as solid lines).
(left) blue$-$green: These are blue points minus the green points from Figure~\ref{fig:delta_z_colors_pwv_los}.
(right) blue$-$red: These are blue points minus the red points from Figure~\ref{fig:delta_z_colors_pwv_los}.
This correlation is reasonably explained by the variation of the calibrated brightness of a M4 star calibrated with a K9 star (orange solid line) through different \pwvlos as shown in Figure~\ref{fig:stellar_pwv_err}.  This model has been set to 0 at the median \pwvlos value of 4.83~mm.
The correlations for each color difference are reasonably explained by a simple linear fit (green dashed line), which reinforces the utility of using \pwvlos as a practical quantity to understand the effect of PWV on absorption.
Note that because this is the $\Delta z_{\rm blue} - \Delta z_{\rm red}$,
negative values here mean that the inferred magnitudes of red stars are fainter than those of blue stars compared to their average magnitudes.  Thus negative values are consistent with increased absorption due to PWV.  This is the same convention as in Figure~\ref{fig:stellar_pwv_err}.
}
\label{fig:delta_z_blue_green_red_pwv_los}
\end{figure}

Comparing the relative trend makes the correlation clearer.
Figure~\ref{fig:delta_z_blue_green_red_pwv_los} clearly shows that there is a dependence of $\Delta z^{\rm blue}_j - \Delta^{\rm green, red}_j$ with PWV across the full range of measured PWV values.
The dependence is well explained by the detailed PWV absorption against stellar templates from Section~\ref{sec:stars}, but also by a simple linear fit to \pwvlos with a dependency of $-0.299$~mmag/mm for blue$-$red, and $-0.114$~mmag/mm for blue$-$green.
Figure~22 in Appendix~B of \citet{Burke18} compares two different DECam exposures, one at a low \pwvzenith and one at a high \pwvzenith.  They find a clear difference in the dependence of $\Delta z$ across 14 bins of stellar $g-i$ color.  Our Figure~\ref{fig:delta_z_colors_pwv_los} only divides the sample into three color bins but shows 32 bins of \pwvlos.

Note that $\Delta z^{\rm blue, green, red}_j$ being 0 at the median \pwvlos of the survey is not dependent on the relative calibration of the blue, green, and red stars in the reference catalog.
We are here looking at the {\em difference} between the per-epoch magnitude and the overall average survey magnitude of the star.
We here clearly detect the slope due to \pwvlos and show that it is explained by variation in the differential PWV absorption of stellar spectral types matching the stellar colors.

We thus have a consistent picture.
Our PWV model explains the zero-point variation from the MzLS survey assuming the images were calibrated against a set of stars with a nominal color of a K9 star.
The color-based residuals for stars of different stellar types are then further consistent with the difference between the effect of PWV on K9 stars and each of the different stellar types.
Specifically, red stars have a median magnitude offset with respect to blue stars consistent with the difference in PWV absorption between M4 stellar spectra and K9 stellar spectra.

\begin{figure}
\plottwo{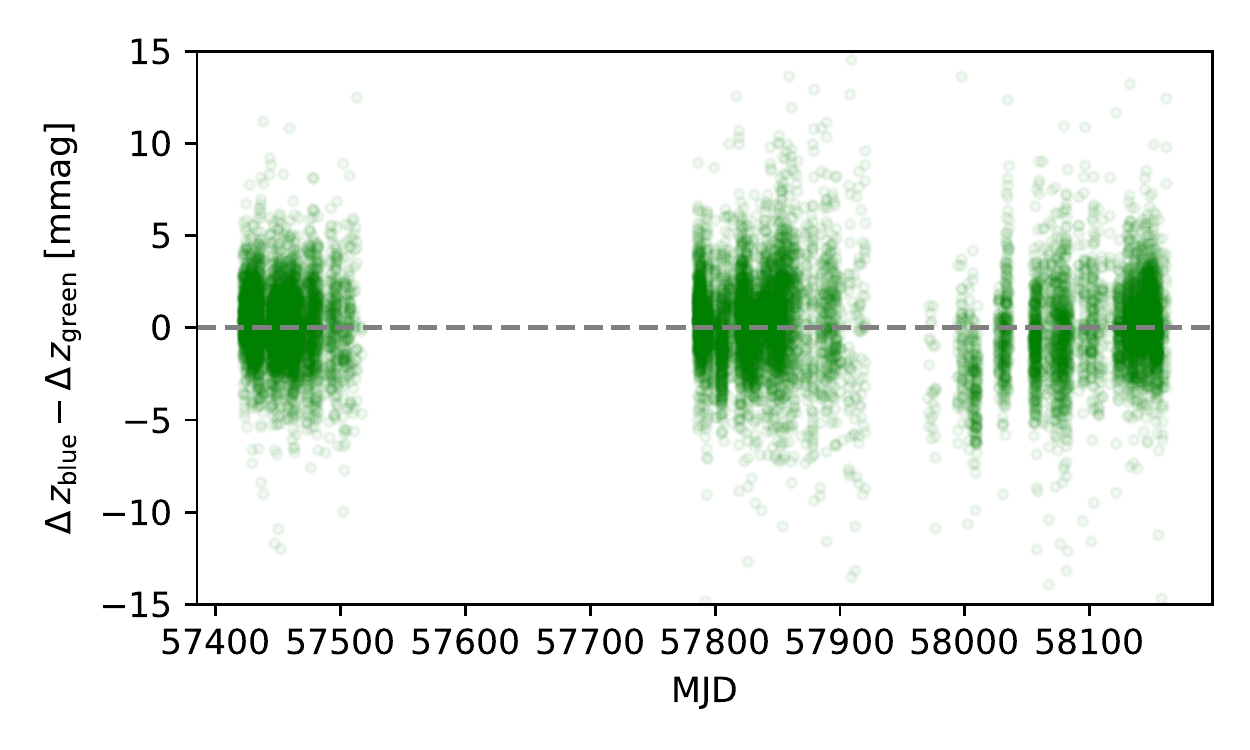}{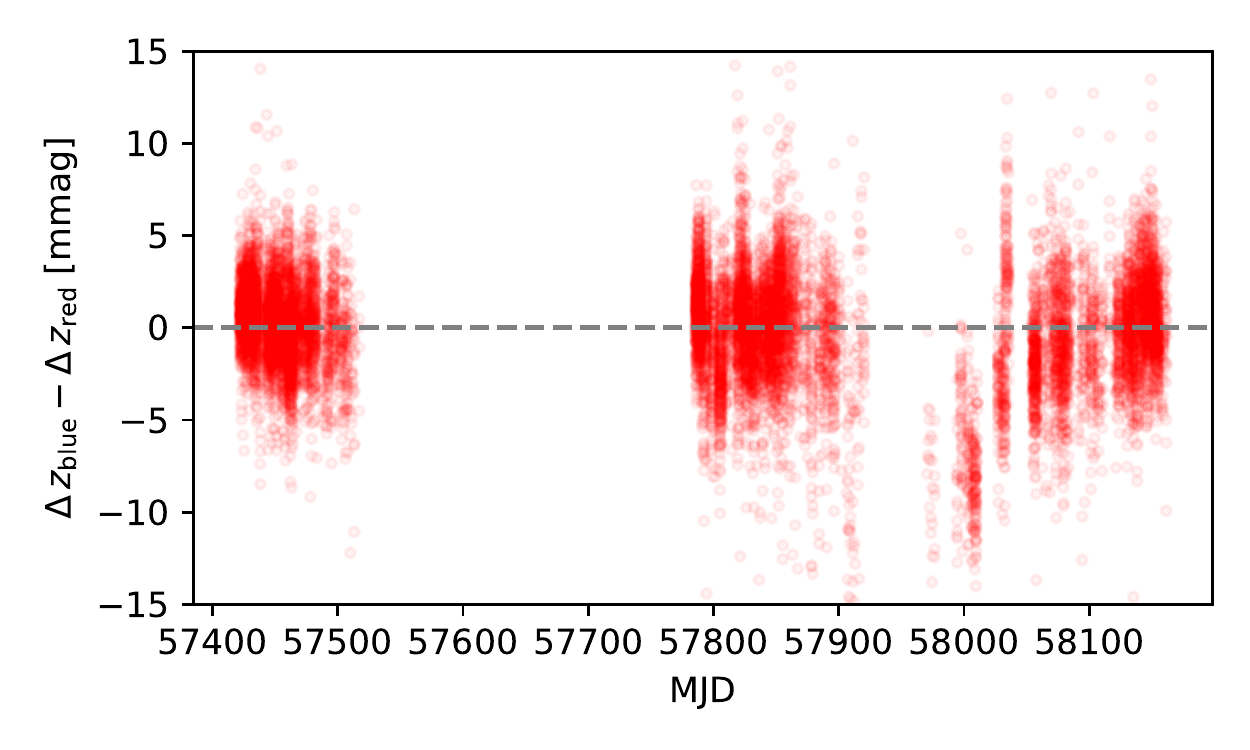}
\caption{
(left) The median blue $\Delta z - $median green $\Delta z$ versus MJD.
(right) The median blue $\Delta z - $median red $\Delta z$ versus MJD.
No correction has been made for PWV.
The only clear trend is the deviation
MJD=57950--58010 which is the time period with the highest PWV ($\sim$25 mm) during the survey.
Note that the secular trend of MJD from Figure~\ref{fig:zpt_mjd_model} should not be apparent in this plot as this is showing the differential measurement as a function of color.
}
\label{fig:delta_z_blue_green_red_pwv_mjd}
\end{figure}

Figure~\ref{fig:delta_z_blue_green_red_pwv_mjd} shows the dependence of
median blue $\Delta z - $ median red $\Delta z$ vs. MJD.
Because there's little correlation between MJD and PWV on these scales, the distribution
should not be expected to show any clear trends.
The deviations are still there, they've just been scrambled up by looking at MJD instead of \pwvlos.
The notable exception is the deviation between MJD 57950--58010.
This was a particularly high PWV period and stands out in both PWV and MJD (see Figure~\label{fig:fig:zpt_pwv_vs_corr_mjd}).
The median blue $\Delta z - $ median red $\Delta z$ provides a separate indication
that the zero-point variation had a strong color term during these nights.

We thus have a consistent picture.  Our PWV model explains the zero-point variation from the MzLS survey assuming the images were calibrated against a set of stars with a median color of a K9 star.  The color-based residuals for stars of different stellar types are then further consistent with the difference between the effect of PWV on K9 stars and those different types.  Specifically red stars have a median magnitude offset with respect to blue stars consistent with the difference in PWV absorption between M4 stellar spectra and K9 stellar spectra.

\section{Correct ZPT with $\Delta z$(color) vs. PWV}  \label{sec:pwv_vs_color}

Section~\ref{sec:delta_z_color} demonstrated the variation in relative brightness of stars of different colors is correlated with PWV.
We here explore whether the PWV data provide additional improvements in ZPT over just using the variation in relative brightness of the stars.  The question is about the basic variation of each measurement, the correlation with ZPT, and the outliers.
An additional important question is whether the color data can supplement cases where the PWV measurements have aberrations.
One could of course ask the question the other way: ``can PWV help identify cases where the color-based differential brightness of stars are incorrectly calibrated?''
But it would be unclear what to then do about those data.
If the color-based differential brightness was systematically in error (rather than just noisy) that would imply there was something wrong in the basic extraction of flux from the image.

\begin{figure}
\plotone{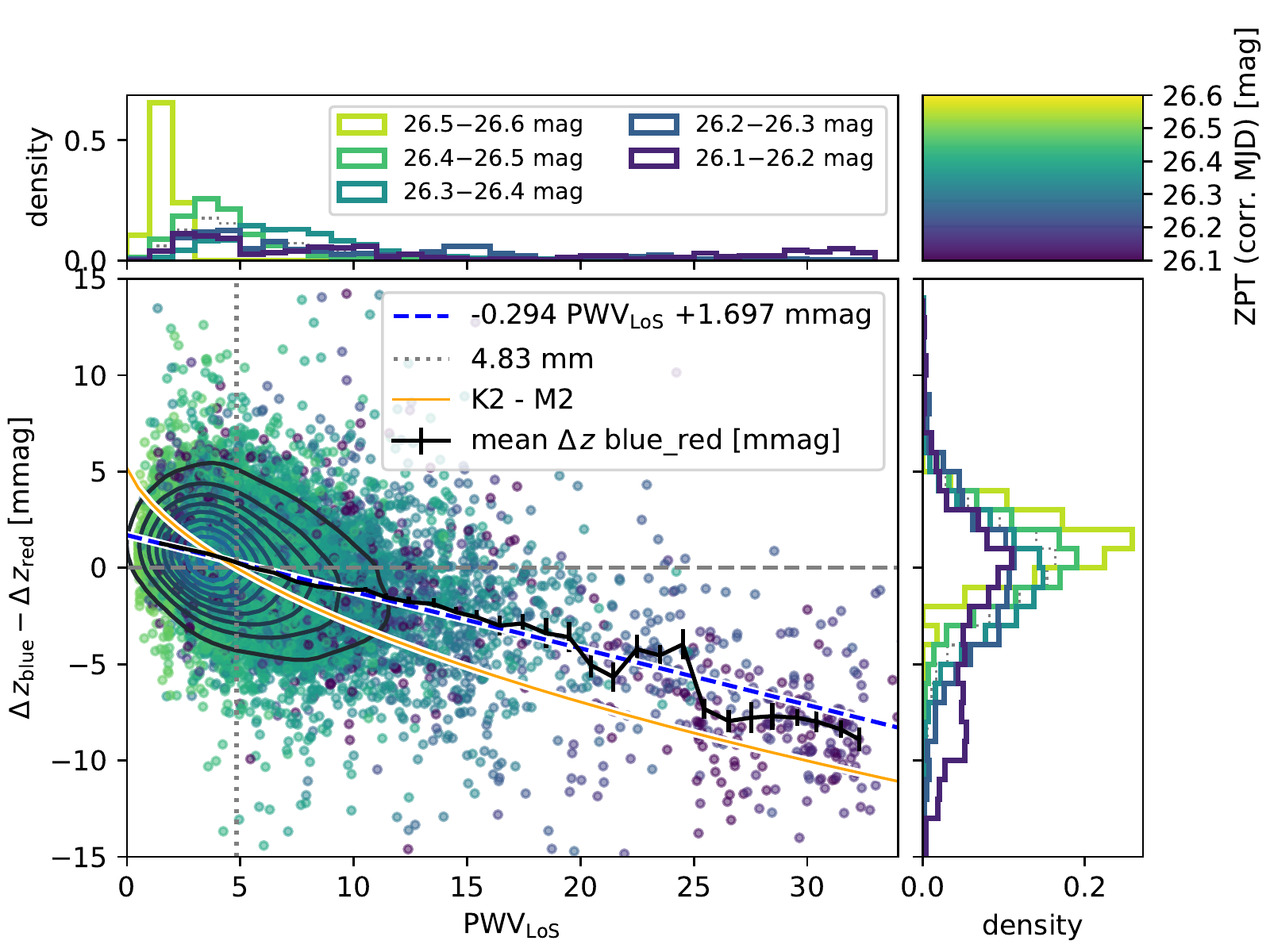}
\caption{\deltazbluered vs. \pwvlos, color-coded by ZPT corrected for MJD trend.  The trend in \deltazbluered vs. \pwvlos is approximately linear in both the mean binned values (black line with errobars) and a simple linear fit (blue dashed line).
A model based on the expected difference in the relative brightness of a M2$-$K2 star qualitatively agrees with the trend (orange line).
The top (side) panels give the projected distributions of \pwvlos (\deltazbluered) for different values of the MJD-trend-corrected ZPT.
The \pwvlos measurements separately much more clearly than the \deltazbluered measurements.
Note that the highest measured values of \pwvlos consistently result in outlier ZPT measurements.}
\label{fig:delta_z_blue_red_pwv_los_zpt_mjd}
\end{figure}

Figure~\ref{fig:delta_z_blue_red_pwv_los_zpt_mjd} shows that \deltazbluered and \pwvlos are clearly correlated, and more negative values of \deltazbluered and higher values of \pwvlos are both associated with less transparent conditions (lower values of MJD-trend-corrected ZPT).
However, the \pwvlos distributions separate much more clearly with respect to MJD-trend-corrected ZPT than do the \deltazbluered measurements.

\begin{figure}
\plotone{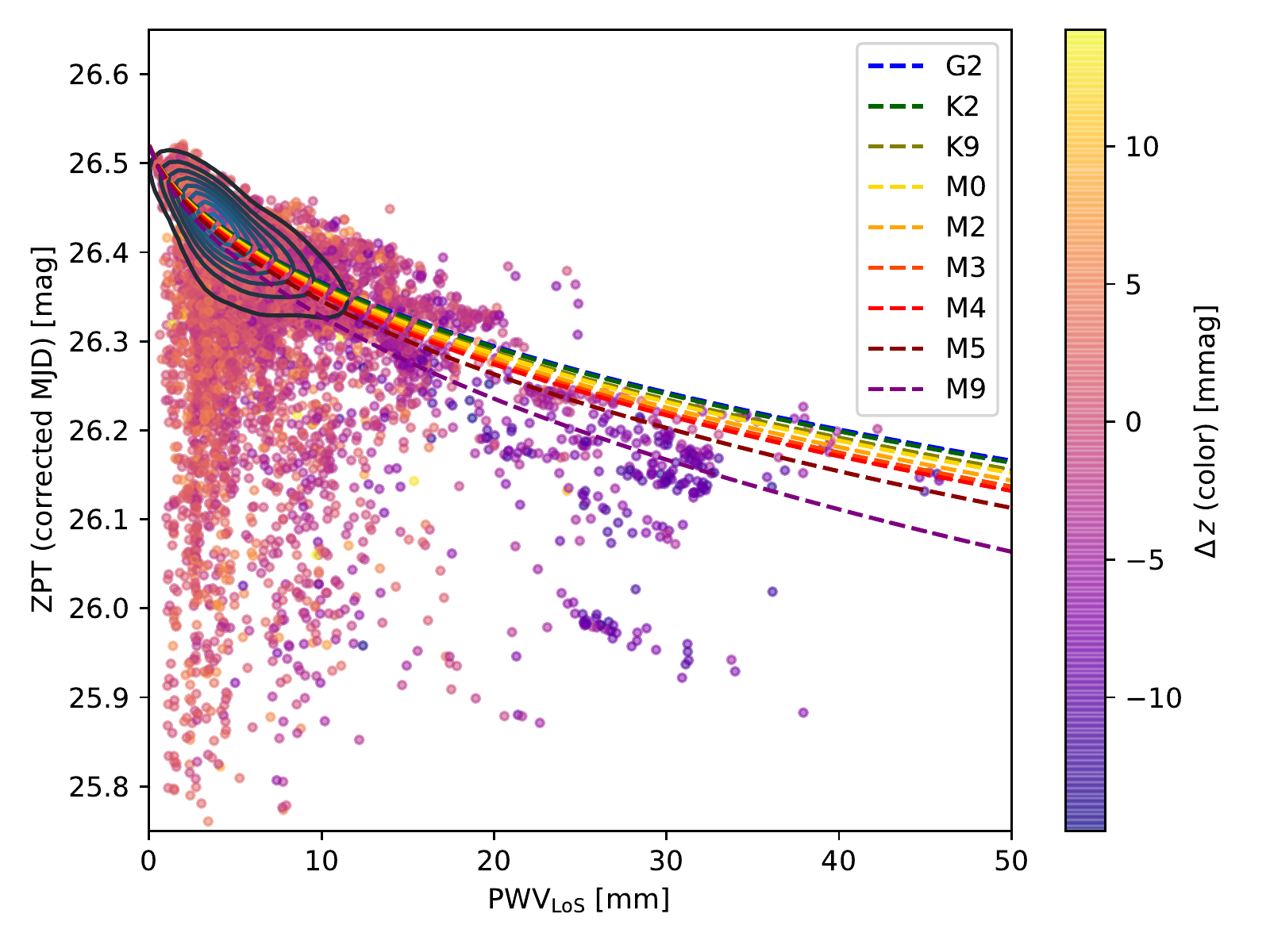}
\caption{
MzLS zero point, corrected for MJD trends, versus \pwvlos, and color-coded by \deltazbluered.
Overlaid are the predicted effect of \pwvlos on six stellar templates: G2 (blue dashed line), K2 (forest green dashed line), K9 (olive dashed line), M2 (orange dashed line), M4 (red dashed line), M9 (purple dashed line).
The points with the largest \pwvlos have clearly reduced $z$-band sensitivity and are also significantly negative in \deltazbluered.
While there is visible population of zero-point variation that is not explained by PWV, these also exhibit small excursions in $\Delta z({\rm color})$.
We thus attribute these to gray extinction variation.
Also note that while those outliers are very visible in the scatter plot, they are in fact a small fraction of the overall population, as indicated by the contour lines of the overall distribution.}
\label{fig:zpt_mjd_corr_pwv_los_hue_delta_z_blue_red}
\end{figure}

Figure~\ref{fig:zpt_mjd_corr_pwv_los_hue_delta_z_blue_red} demonstrates the success of using \pwvlos to predict MJD-trend-corrected ZPT.
The expected relationship across several stellar types explains both the core of the MJD-trend-corrected ZPT variation as well as the outliers.
There are a noticeable number of outliers that have lower values of MJD-trend-corrected ZPT even at low values of \pwvlos.  However, these images also have low \deltazbluered (refer back to Figure~\ref{fig:delta_z_blue_red_pwv_los_zpt_mjd}) and we hypothesize that these images have some amount of gray extinction.
The residual of MJD-trend-corrected ZPT vs. the predicted stellar model dependence on \pwvlos
do not appear to be obviously correlated with \deltazbluered.

\begin{figure}
\plotone{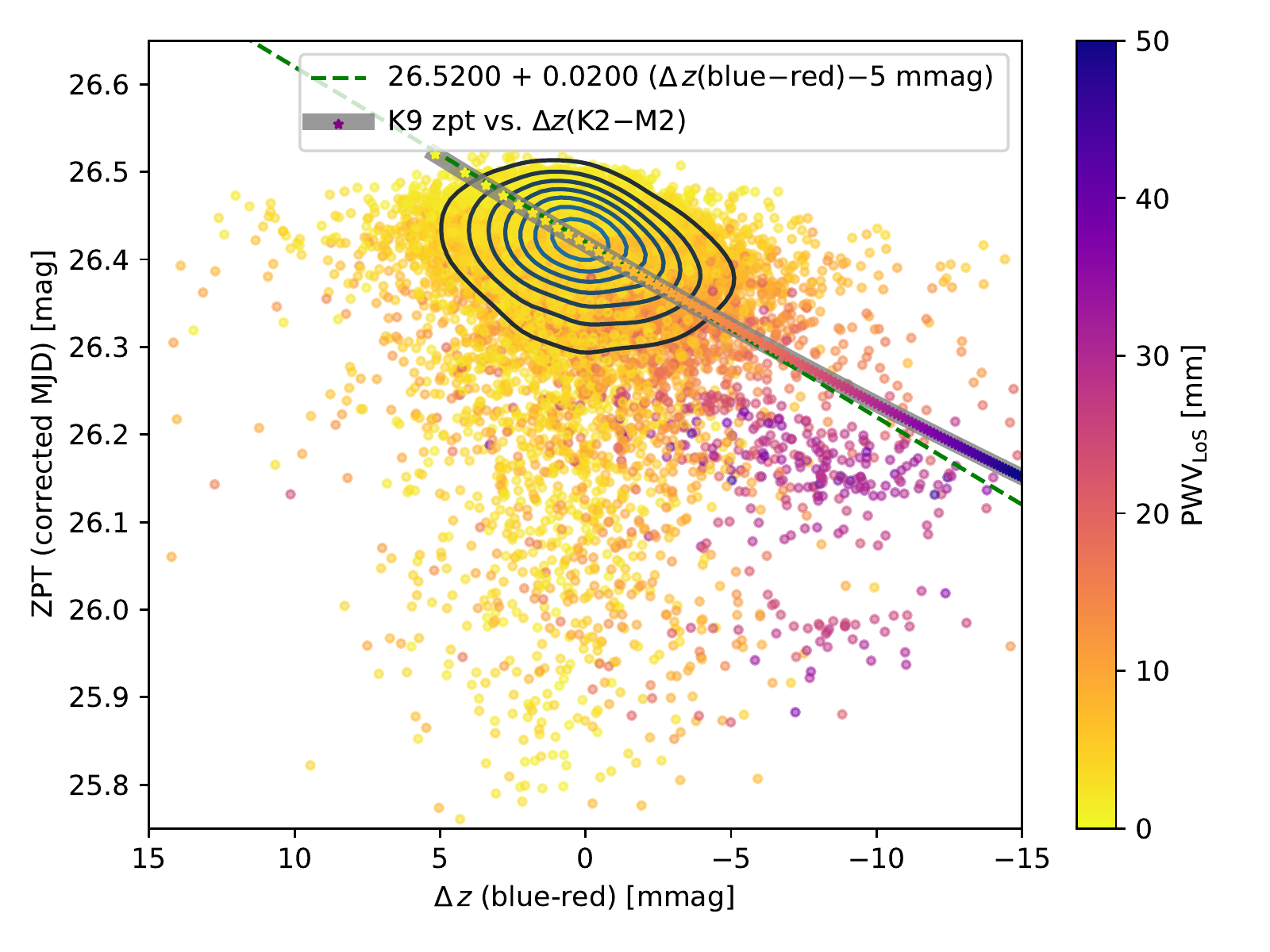}
\caption{
MzLS zero point, corrected for MJD trends, versus \deltazbluered, color-coded by \pwvlos.
(green dashed line) A simple linear relationship explains the major axis of the contours.
(grey line + stars color-coded by \pwvlos) A model of the expected zero point change for a calibration based on a K9 star
versus the expected change in relative apparent brightness between K2$-$M2 star.
The model curves up from the linear model as more water lines saturate -- the M2 SED remains more sensitive to the increased PWV than the K2 or K9 SEDs.  While the observed zero point is in agreement with the stellar model (project the model PWV colors and the data colors onto the zero point axis), the observed \deltazbluered data don't go as negative as expected by the model.
}
\label{fig:zpt_mjd_corr_delta_z_blue_red_hue_pwv_los}
\end{figure}

Figure~\ref{fig:zpt_mjd_corr_delta_z_blue_red_hue_pwv_los} shows that \deltazbluered is indeed correlated with MJD-trend-corrected ZPT, but is not as powerful at predicting the ZPT variation, particularly at the highest level of PWV.

We conclude that \pwvlos is a significantly better predictor of MJD-trend-corrected ZPT than \deltazbluered.
Measurements of \pwvlos can improve the standard deviation of the measured MJD-trend-corrected ZPT from 58~mmag to 30~mmag.  Using, an admittedly simple, linear model of \deltazbluered to attempt to improve the MJD-trend-corrected ZPT variance results in a worse standard deviation of 69~mmag.

We end this section by mentioning two opportunities to improve the differential brightness color-based measurement of PWV:
(1) survey with more repeated observations of fields in more filters;
and (2) taking advantage of the temporal and spatial correlation of PWV.

MzLS was a survey conducted in a single band with a typical limit of only three repeated observations of a star.
It's possible that results for color-dependent $\Delta z$ could potentially be improved in surveys that have more repeated observations over different conditions of the same field, or taking several images in different filters within an $\sim$hour to capture more directly the effects of PWV versus gray extinction.
The gray extinction variation can be dominant over the effect of PWV.
Thus measuring the gray extinction separately might allow for a cleaner differential brightness color-based measurement of the effect of PWV.

Greater power may come from including the strong temporal and spatial correlation of PWV.
The model of a uniform slab across the sky varying on 30--60 minute time scales is remarkably good.
It's possible that assuming some time constant for variation of PWV would allow the combination of measurements across a set of images to improve the sensitivity of using the differential brightness color-based method to correctly predict the true transmission spectrum.

\section{Impact on Selected Science Cases} \label{sec:science_cases}
We have established that PWV explains the non-secular zero-point variation of the MzLS survey
and that there is a color-dependent effect for the inferred magnitudes of stars of different colors and spectral types.  We next explore the consequences for PWV absorption for a wider variety of objects.
We calculate the expected induced photometric error for a variety of object SEDs grouped into stars, supernova, and quasars.

\subsection{Stars}\label{sec:stars}
\begin{figure}
\centering
\includegraphics[width=2.3in]{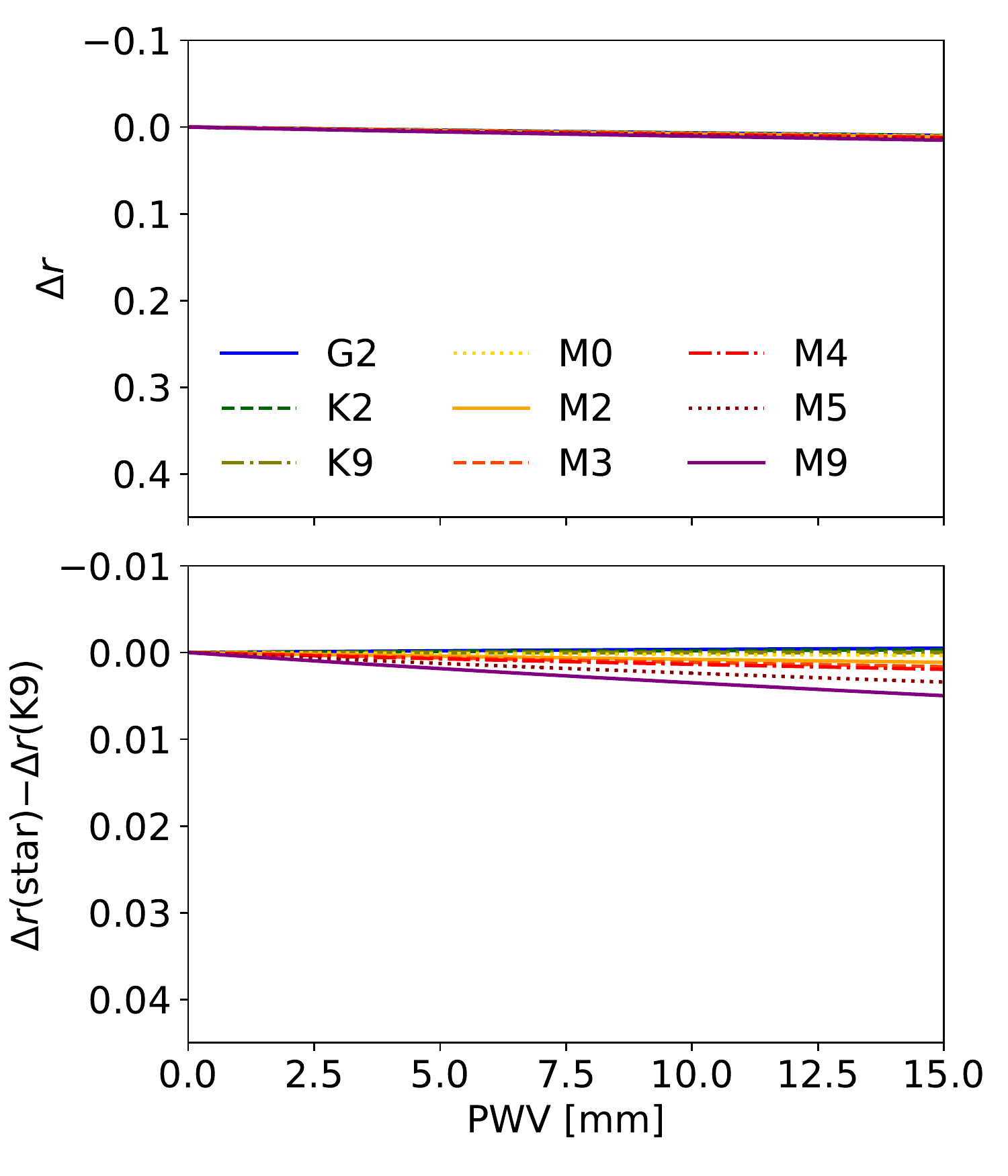}
\includegraphics[width=2.3in]{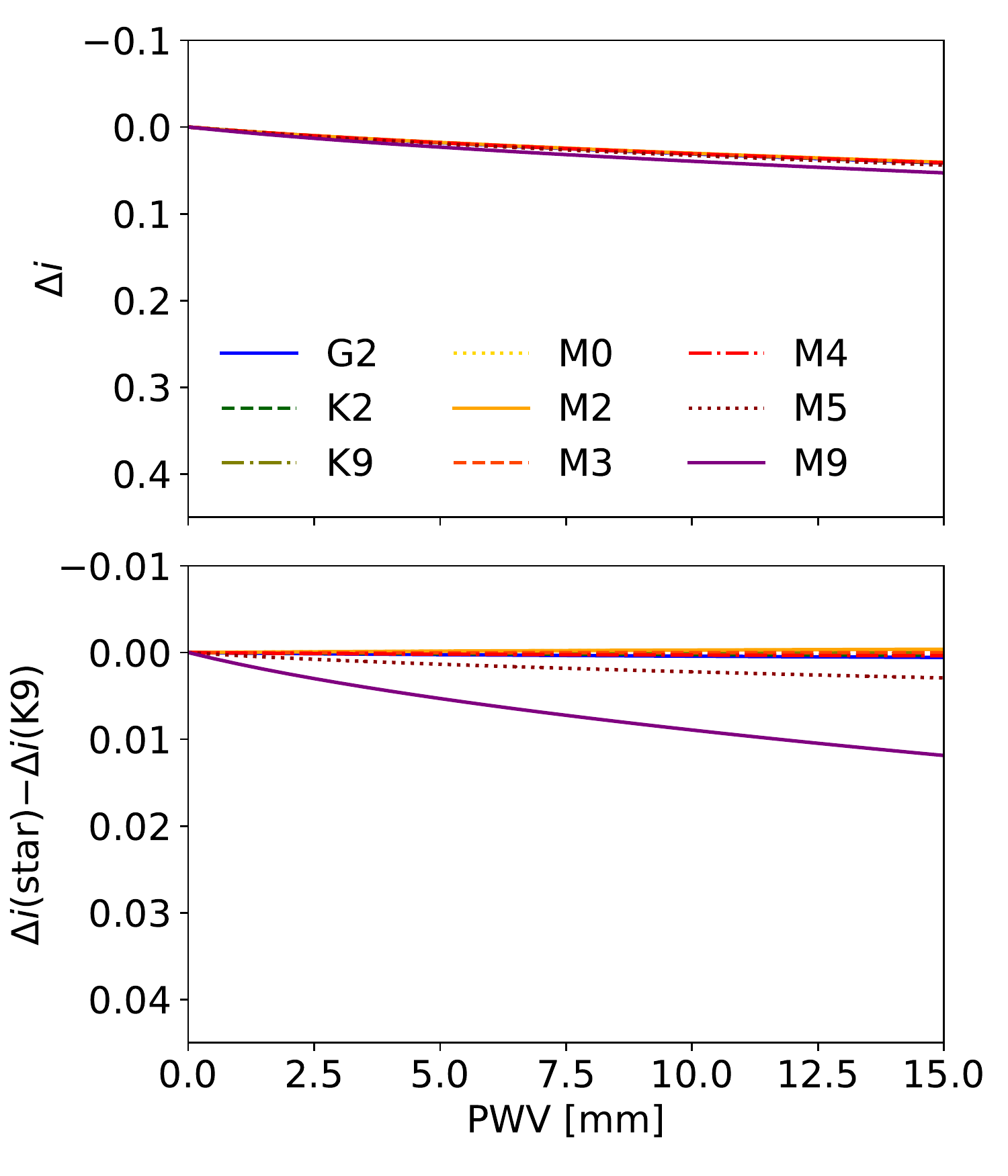}
\includegraphics[width=2.3in]{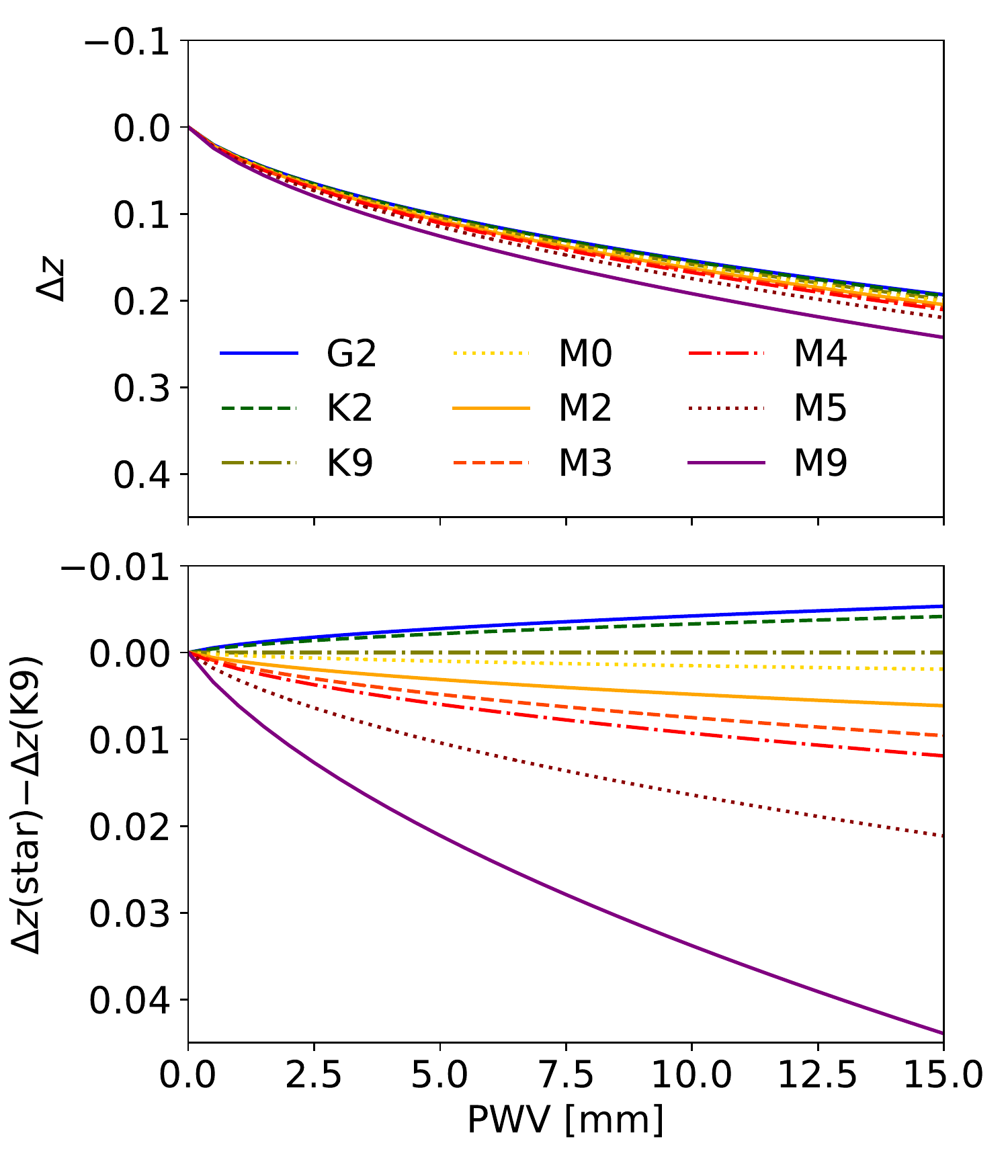}
\caption{
Change in stellar $r$-band (left), $i$-band (center), and $z$-band (right) photometry for G2, K2, K9, M0, M2, M3, M4, M5, and M9 stars (top) and the induced error from PWV assuming target$-$reference star combinations (bottom).
The magnitude scale for the top panels is 10 times that of the bottom panels.
}
\label{fig:stellar_pwv_err}
\end{figure}

Achieving both accuracy and precision in ground-based stellar photometry requires a full understanding
of the system transmission function, including the variable atmosphere. In particular,
objects whose SEDs differ from the reference catalog stars are affected differently by variable PWV absorption.
Color terms can handle smooth differences in SEDs, which is particularly effective when looking at the Rayleigh-Jeans tail of stars.
However, cooler stars, particularly M, L, and T stars (a) peak in emission at NIR wavelengths; and (b) have very non-blackbody SEDs with complex molecular band features.
The impact of second order telluric absorption effects interferes with achieving precise photometry for exoplanet transit searches \citep[e.g.,][]{Baker17}.
Accurate colors are additionally important for reliable object classification.

To demonstrate the effect that PWV absorption has on stellar photometry, we integrate PHOENIX models~\citep{Husser2013} multiplied by a TAPAS water vapor transmission spectrum scaled to different PWV values.
These spectra are then integrated over filter profiles to compute the flux in each band as a function of PWV.
These fluxes are converted to magnitudes in each band, $m$, referenced to a PWV of 0~mm such that $\Delta m = m(\pwv) - m(\pwv=0)$.
In Figure~\ref{fig:stellar_pwv_err} we show $\Delta m$ for the $r$, $i$, and $z$ passbands in the top panels of each figure for K9, M4, and M9 spectral types.

\begin{figure}
\plotone{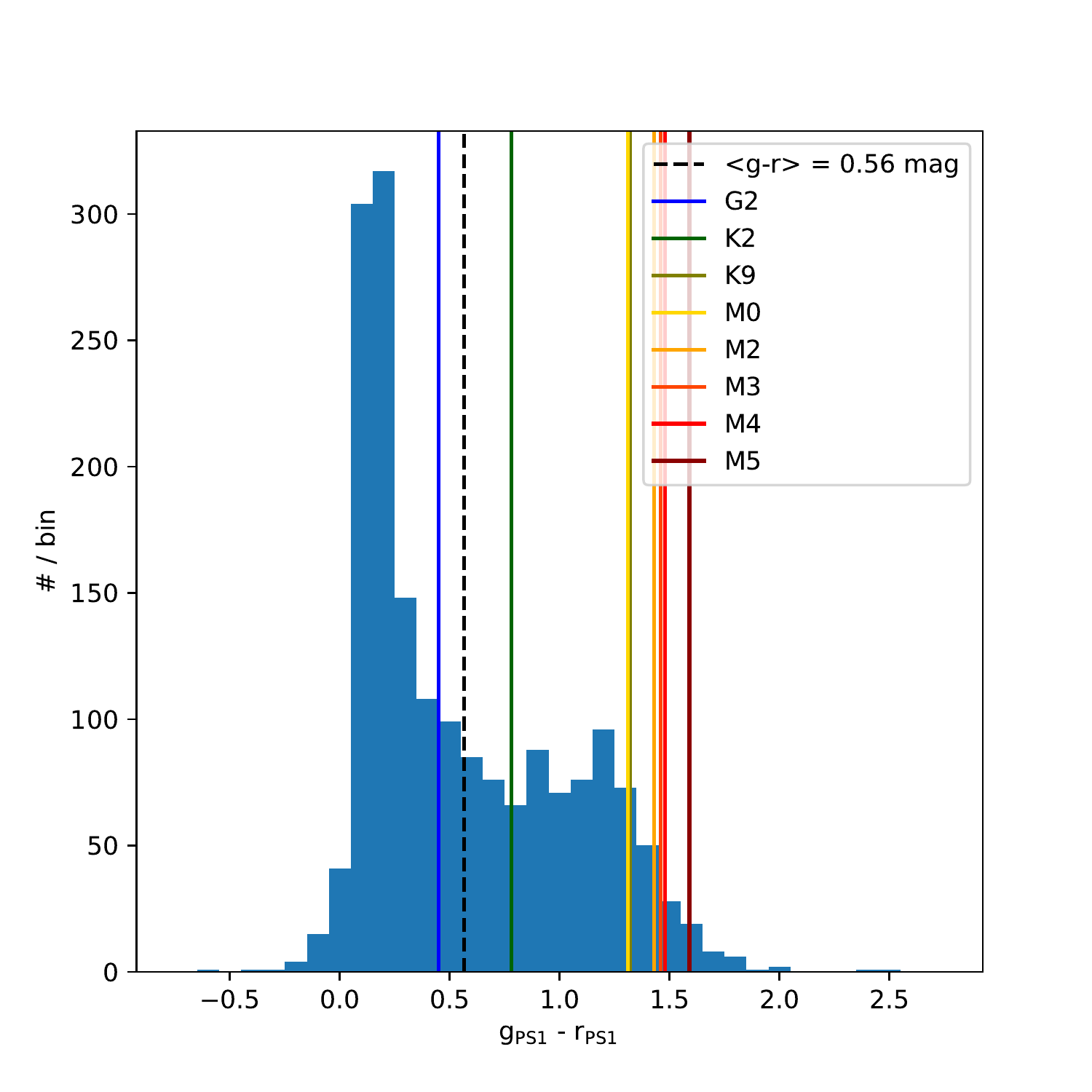}
\caption{The $g-r$ color distribution of the PS1 reference catalog used for the MzLS calibration.  Reference stars of different colors will be affected differently by varying PWV amounts.  The mean color is between that of a K9 star and a K2 star.  If one assumes that the calibration effectively takes an average over the mean color of the reference catalog, then the PWV effect should go approximately with the mean color of the reference stars.}
\label{fig:ps1_ref_cat_color}
\end{figure}

In the MzLS catalog, the typical reference star used for calibration is a K9 dwarf (see Figure~\ref{fig:ps1_ref_cat_color}).  A zero point estimated for these reference stars does not correctly include the effects of PWV absorption on the integrated photometry of redder stars. The bottom panels in Figure \ref{fig:stellar_pwv_err} show this expected zero point error by plotting the difference in the effect of PWV absorption for K9, M4, and M9 stars.
In nominal observing conditions (${\rm PWV}=5$~mm), the error in the $z$-band calibrated brightness of M4 and M9 stars is 8 and 25~mmag, respectively, when calibrating with a K9 star.
If instead an M4 star is used as a reference in calibration, an M9 star still has a significant brightness error of 17~mmag.
This comparison shows how sensitive the effect is for stars with SEDs that vary quite substantially over a seemingly small temperature range. In the $i$ band, the differential extinction between M4 and K9 stars is near zero whereas an M9 star would still experience a 5~mmag error. Due to the minimal amount of absorption in the $r$ band, the error in this band is below 2~mmag for each star.

The $i-z$ color error due to these incorrect zero point corrections for an M9 star are: 15~mmag bluer at \pwvlos=5~mm and $\sim$30~mmag bluer at \pwvlos=10~mm.
Of importance to time series data is how the magnitudes would change in time as \pwvlos changes between observations.
For this, the slopes of the curves in Figure \ref{fig:stellar_pwv_err} show that the $z$-band magnitude for M4 and M9 stars referenced to a K9 star will change approximately 2 and 5~mmag per millimeter change in PWV, respectively, in drier conditions typical of conditions at KPNO (and Cerro Pach\'on).

\subsection{Supernova Cosmology} \label{sec:supernovae}

The use of Type Ia supernovae (SNe~Ia) as cosmological probes relies on them having homogeneous light-curves with standardizable luminosities at the time of peak brightness. By calibrating for intrinsic variations between the absolute magnitudes of individual SNe~Ia, cosmological models are used to fit SN~Ia distance moduli ($\mu$) as a function of redshift \citep[for example, see][]{Riess98, Perlmutter99, Betoule14, Scolnic18, Abbott19}. However, as SN~Ia samples grow larger, the variation in brightness between individual SNe no longer becomes the limiting issue. Instead, common systematics that change the effective average difference for large subsets of the sample become important.
Photometric calibration, and a proper treatment of PWV, are thus key in achieving the goals of large surveys such as LSST.

The importance of calibration in controlling cosmological uncertainties is well demonstrated in the literature.
In the recent cosmological analysis of the Pantheon SN~Ia sample~\citep{Scolnic18}, photometric calibration uncertainties contributed between 2 and 6~mmag, constituting up to half of the total uncertainty in cosmological parameters.
Similarly, in \citet{Lasker19} system transmission uncertainties introduced a redshift-dependent bias in SN~Ia luminosity distances of $\sim0.02$~mag.
While we note in both cases that atmospheric effects were not considered separately from instrument throughput,
we argue that variable atmospheric absorption can play a significant role in the $z$-band calibration uncertainties, particularly in thicker CCDs with increased sensitivity in the 940--980~nm water band.

The SEDs of SNe~Ia are quantitatively different than stars.  They have wide absorption lines characteristic of the elemental composition and 10,000~km~$s^{-1}$ typical explosion speeds.
SN~Ia light-curve fitting uses this and models the spectrum of SN~Ia as part of the forward modeling, but it needs an accurate and precise transmission function to do so.
Photometric calibration based on the stars in the image traditionally provides just a color term, which is the equivalent of a smooth variation of the transmission function.
To precisely calibrated SNeIa, the full transmission function needs to be determined, including the highly non-smooth variation due to PWV absorption.

\begin{figure*}
    \centering
    \includegraphics[width=0.5\textwidth]{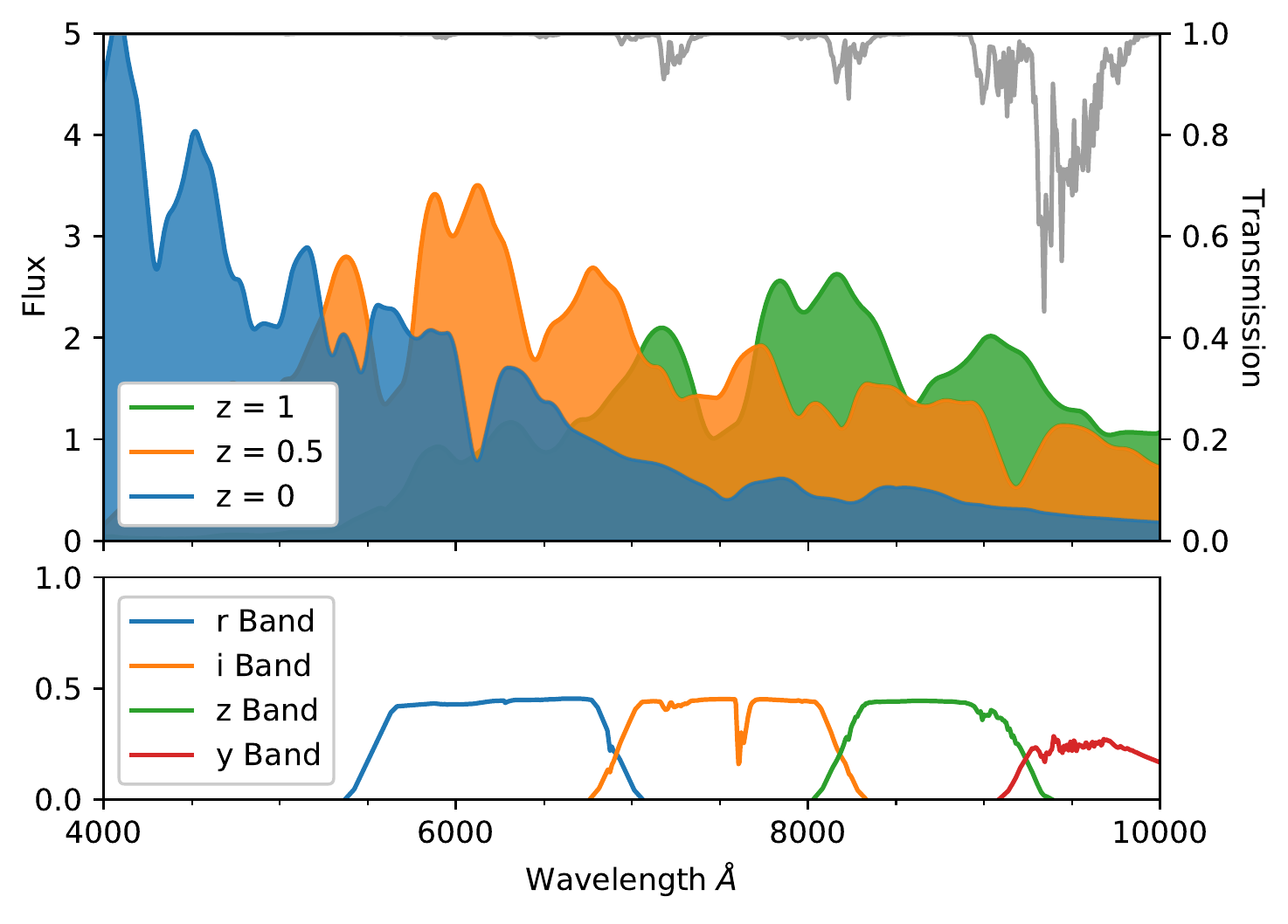}
	\caption{
    The spectral template of the Salt2 supernova model for a redshift of 0.1 (blue), 0.5 (orange), and 1 (green) against the PWV transmission (grey) for a PWV concentration of 4 mm.
    SN~Ia spectra are dominated by several broad absorption features, many of which have equivalent-widths similar to the widths of PWV absorption features in the $zy$ bands.
    As the template spectrum is shifted to higher redshifts, different spectral features begin to overlap with the strong PWV absorption feature at $\lambda \approx 950$~nm.}
    \label{fig:sn_spectrum}
\end{figure*}

The complex structure of PWV absorption features is neither uniform nor constant.  The variation in PWV absorption causes the apparent brightness of observed targets to vary as a function of redshift, as different features are redshifted across the water absorption bands.
This effect is most significant for high-redshift SNe~Ia ($z \sim 1$) when the brightest region of the SED ($\lambda \approx 400$~nm) begins to enter the $z$ and $y$ bands (see Figure~\ref{fig:sn_spectrum}).
However, even for low redshift targets, temporal variations in the atmospheric transmission function introduce a bias in the estimated color of SNe~Ia.

To demonstrate how PWV absorption impacts photometric SNe~Ia observations, we use the SALT2 spectroscopic template from \cite{Guy07} to simulate sets of SN~Ia light-curves
for a range of PWV values and over a redshift range of $0 \lesssim z \leq 1$.
We assume a fixed PWV concentration for each set of SNe~Ia.
The simulated light-curves are then fit using the SALT2 model without a PWV component.
For simplicity, we simulate SNe~Ia having the fiducial model stretch ($x_1 = 1$) and color ($c = 0$).  We also fix $t_0$ and $z$ to the simulated values.
We simulate a daily observational cadence in each of the $rizy$ bands.
The supernova is thus sampled in phase from (-15 -- +50)/(1+z) days.
The other SALT2 model parameters, $x_0$, $x_1$ and $c$, are left free with flat, independent priors.

\begin{figure*}
    \plotone{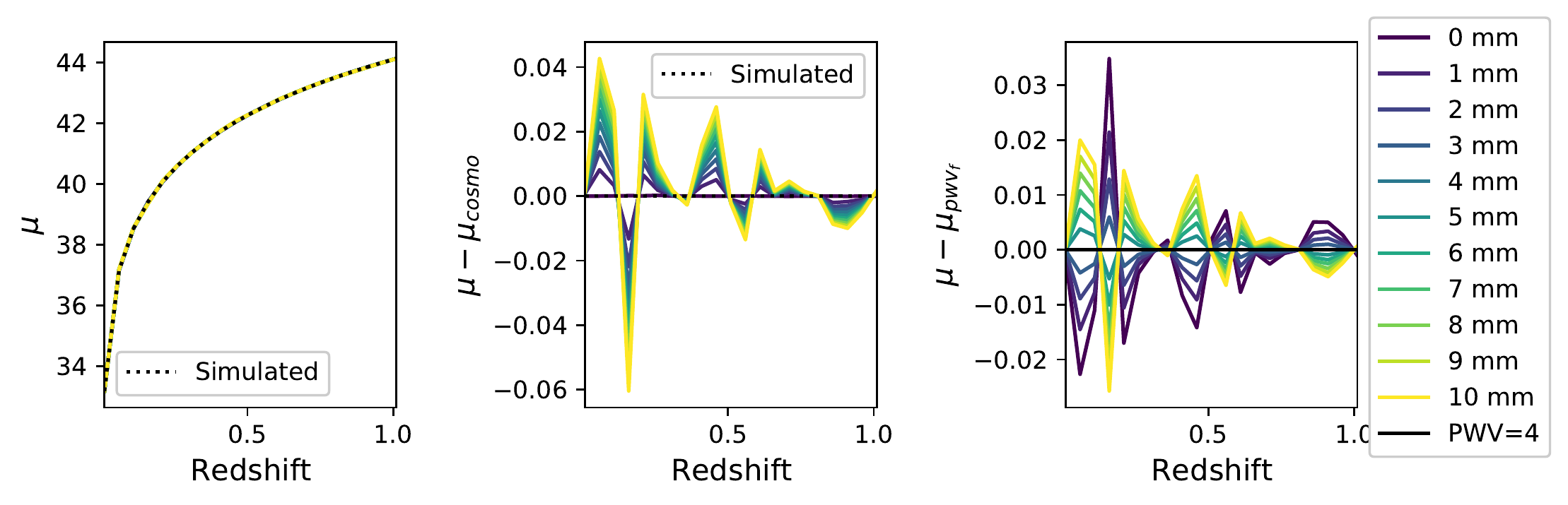}
	\caption{Distance modulus residuals ($\Delta\mu$) determined from a set of simulated SNe~Ia, each suffering from a different level of PWV absorption (assumed constant for the full light curve). Residuals are shown both relative to the underlying cosmology of the simulation (center) and to a fiducial atmosphere of 4~mm PWV (right).
    The pattern with redshift comes from the broad features of SN~Ia spectra moving across the water vapor absorption features.
    }
    \label{fig:sn_mu}
\end{figure*}

Figure \ref{fig:sn_mu} shows the residuals in the distance moduli as determined from the simulated SNe.
Note that this makes the unphysical assumption that PWV is constant for all points in the SN~Ia lightcurve.
In realistic conditions, PWV would vary and these biases would be averaged down.
However, a season dependence of average PWV would translate into a bias as a function of RA
as there is a correlation between average PWV and observed RA.
Even for relatively dry, photometric conditions (PWV $\lesssim 4$), uncorrected PWV absorption introduces a bias of up to $0.02$~mag in the estimated $\mu$.
A similarly sized shift is also found for low redshift targets suffering from PWV absorption that varies by more than a few mm away from the assumed fiducial atmospheric value.
This is just another reminder that the assumption of a fiducial transmission model that is constant from image to image, such as when constructing the effective bandpass throughputs for a given survey, is not sufficient for achieving the millimag level of precision anticipated from future large scale surveys.

We note that our results are similar in size to Hubble residuals found in existing cosmological analyses.
For example, after assuming a mean atmospheric absorption model to calibrate instrumental magnitudes, the cosmological analysis of \cite{Betoule14} found average residuals in the fitted $\mu$ of approximately 0.01 and 0.06~mag for their low and high redshift SN samples respectively.
Calibration uncertainties were the largest contributing factor in measurement uncertainty, and made up over 36\% of the measurement uncertainty in $\Omega_m$.

\subsection{Quasars} \label{sec:quasars}

\begin{figure}
    \plotone{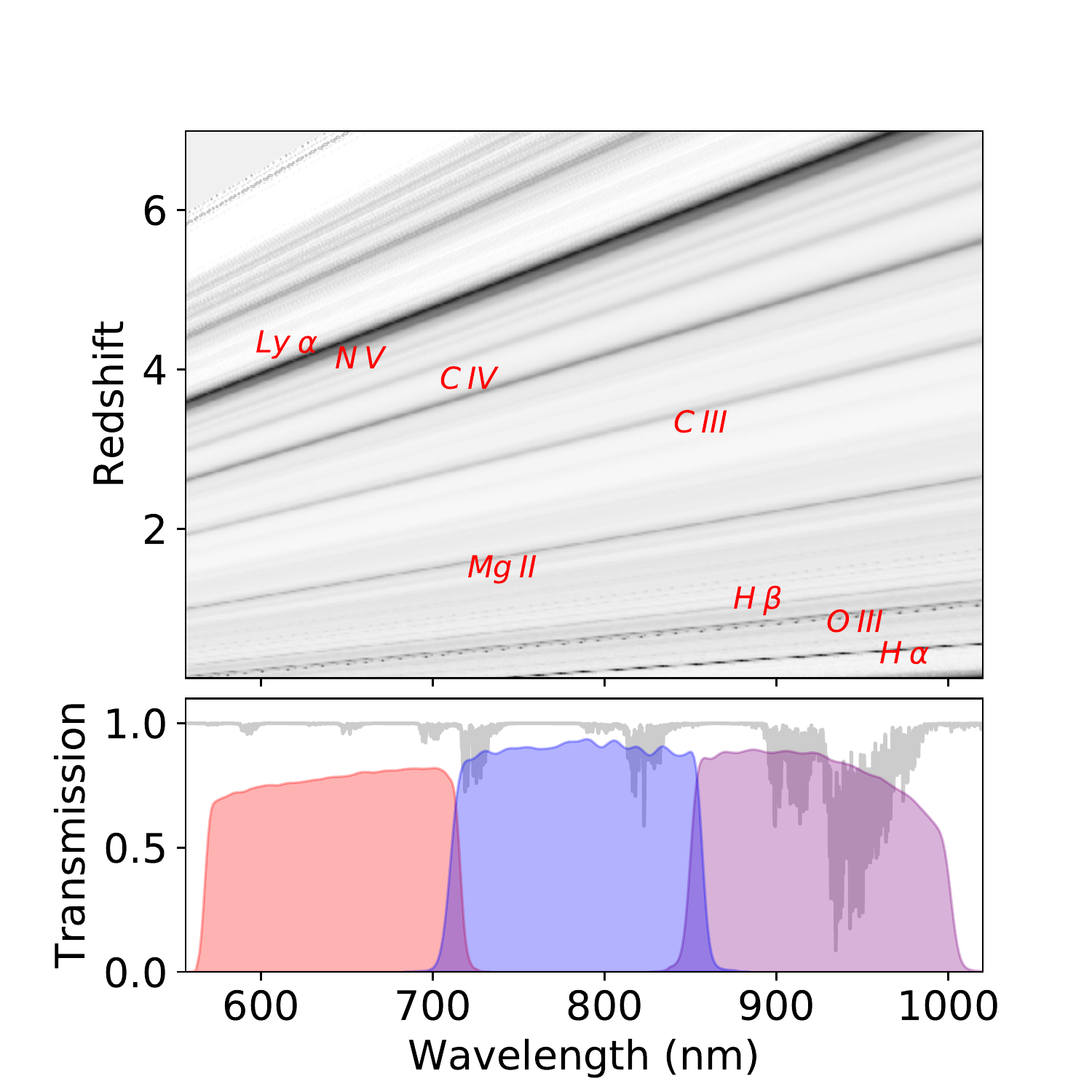}
	\caption{(top) Quasar spectra at different redshifts with prominent emission lines and (bottom) telluric spectra and r, i, and z MzLS filter profiles.}
	\label{fig:quas_setup}
\end{figure}

\begin{figure}
	\plottwo{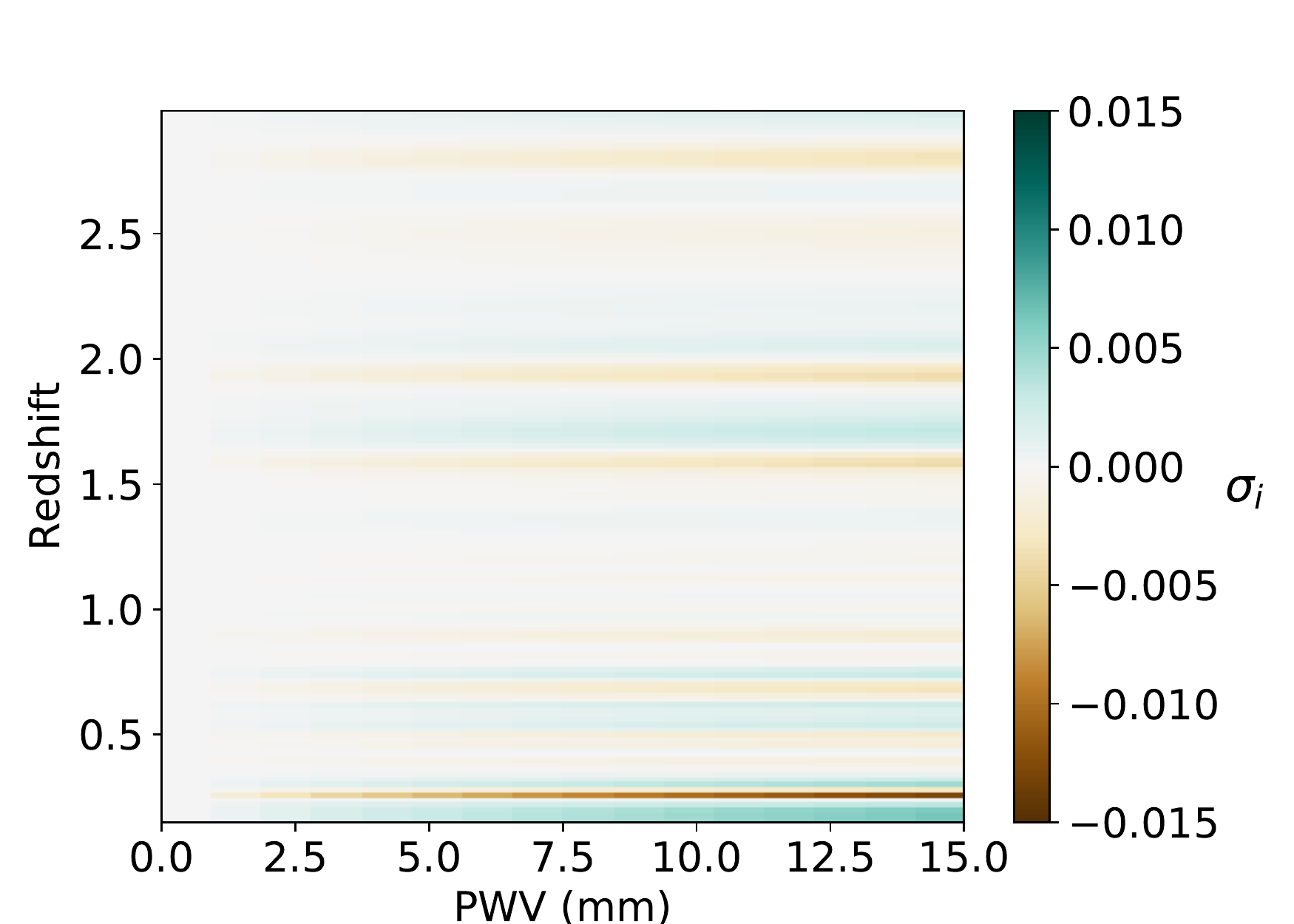}{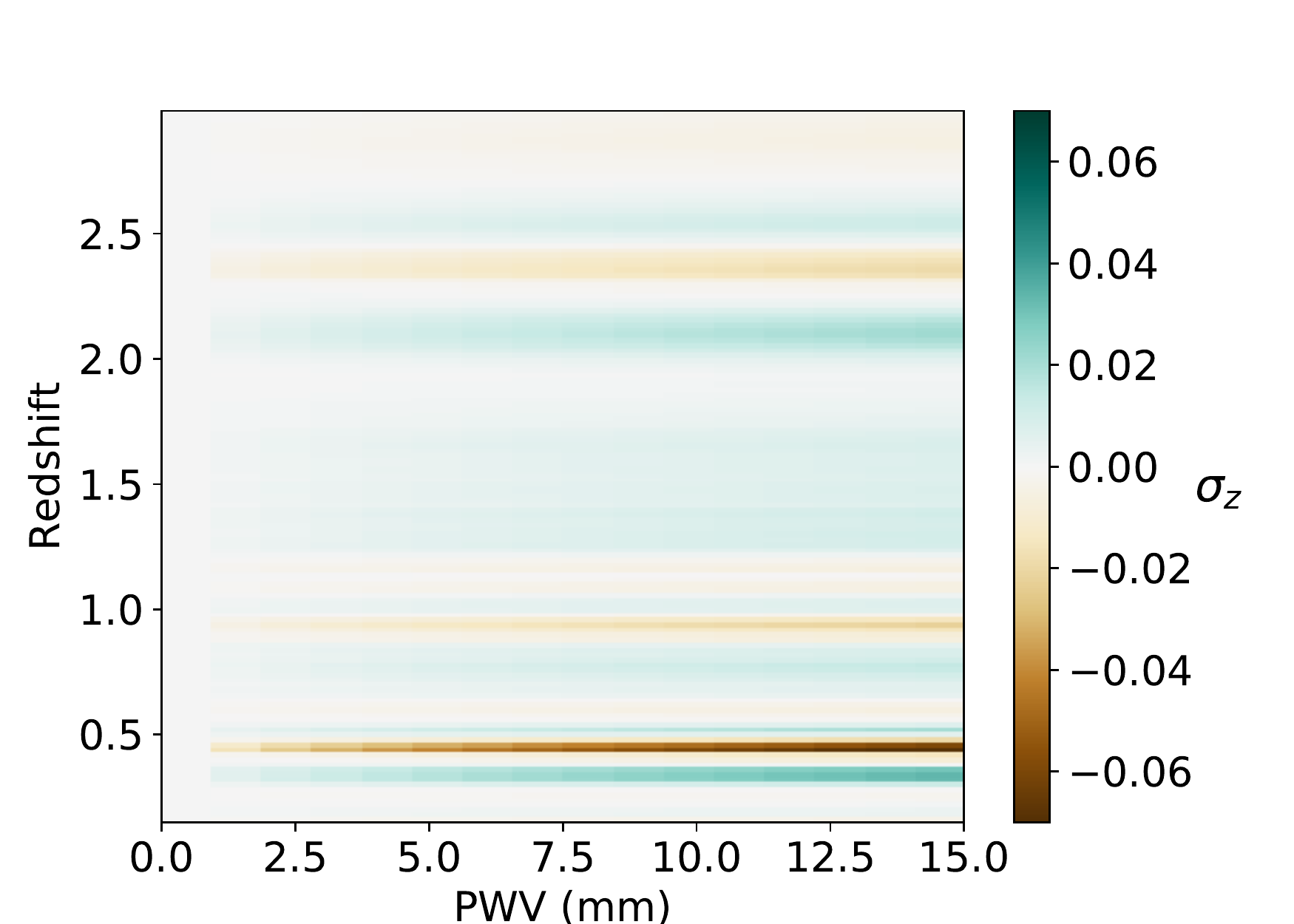}
	\caption{Error in the i band (left) and z band (right) of a composite quasar at various redshifts and PWV values.}
	\label{fig:quas_err}
\end{figure}

The sample of known quasars has grown substantially in last two decades thanks to surveys like SDSS, Pan-STARRS, and DES \citep{Richards2009, Schindler_2019, Tie2017}. The current sample of over a million quasars is expected to grow tenfold with LSST \citep{Abell2009}. With such a large sample, obtaining spectra will not be possible for all objects and photometrically derived redshifts will therefore remain important classifiers. Furthermore, sampling the temporal variability of quasars photometrically enables the study of the physical nature of these objects \citep{Schmidt_2011}.

The spectral energy distribution of a typical quasar is marked by several prominent emission features including Lyman~$\alpha$ in the UV and H$\alpha$ emission at redder wavelengths.
In Figure~\ref{fig:quas_setup} we show how these and other prominent emission features shift with redshift causing them to overlap with absorption features due to water vapor in Earth's atmosphere.
Due to the large flux contribution by these emission features in a given photometric band, when the features align with Earth's atmospheric absorption there is the potential for a significant error in the zero point correction that will be a function of the PWV during the observation.

To quantify the error in the photometry of a typical quasar spectrum, we use the composite quasar spectrum described in \cite{Vanden_Berk_2001}. This spectrum was generated by averaging together over 2000 SDSS spectra to provide a high SNR spectrum that is characteristic of a typical quasar spectrum and covers a wide restframe wavelength range of 80--855~nm. Using the TAPAS telluric transmission spectrum and this composite spectrum, we compute the change in magnitude in $r$, $i$, and $z$ passbands and compare it to the extinction in a K9 reference star in the same atmospheric conditions.  We compute this error term on a redshift grid ranging from 0.15 to 3.
The error in $r$ is less than 1~mmag at all values of redshift and PWV, while the $i$ and $z$ bands have errors as high as 15 and 60~mmag, respectively, for the highest PWV of 15~mm. Figure \ref{fig:quas_err} shows the full effect for $\sigma_i$ and $\sigma_z$ in grids of redshift and PWV. As expected, the largest errors are observed when spectral features overlap with absorption in Earth's atmosphere. Quasars at $z~0.4$ have the highest error due to H$\alpha$ overlapping the strong 940~nm water vapor band.

Quasars vary in brightness on time scales from hours to years~\citep{Ulrich1997,VandenBerk2004,deVries2005,Kimura2020}.
While 20--60~mmag changes in $z$ are quite small with respect to the long-term intrinsic variability of quasars, on short time scales, the PWV-induced variability can be greater than the short-term intrinsic variability of quasars.
If the cadence of observations is high enough and the PWV during observations is known, this absorption could help constrain the redshift of an object.
Other efforts to utilize the effects of Earth's atmosphere to aid photometric redshift estimates include the works by \cite{Graham18} and \cite{Kaczmarczik09}.
In \cite{Kaczmarczik09}, the authors show how atmospheric refraction causes objects with different SEDs to experience different positional offsets, which can help break the degeneracies in photometric redshift estimates.
The authors therefore suggest that surveys like LSST observe several frames of a part of the sky at high airmass to increase the magnitude of the differential chromatic refraction between survey objects.
\cite{Graham18} also show how higher airmass observations can produce changes in an object's magnitude that is a function of the object type.
Since water vapor absorption is variable in time, looking for correlations between the $z$ magnitudes (or LSST $y$) with PWV in time would further aid in breaking degeneracies.

The PWV-induced errors in quasar color may also be pertinent to studies monitoring the color variability of these objects.
This added scatter in the $r$ and $z$ bands should be noted in such an analysis for quasars at redshifts that maximize the color error.
Because quasars are prevalent throughout the sky, another use of this effect could be to generate a sample of quasars at redshifts leading to overlap with prominent spectral features that can then be used to track changes in PWV.
For this to work, many quasars would need to be monitored to average down their intrinsic variability in order to reveal their common mode variability due to changing PWV.
Wide-field surveys such as LSST may observe sufficient quasars in one exposure to identify the common-mode variability.

\section{Discussion and Conclusions} \label{sec:conclusions}

The dominant effect in zero-point variation in MzLS was a long-term secular trend that was not correlated with either the GPS-measured PWV or with the observed color-dependent relative brightness of observed stars.
This secular trend was thus consistent with uniform loss of sensitivity across the $z$ passband.
We use independently measured PWV values through a dual-band GPS system was to successfully model much of the remaining variation.

In principle, the effect of PWV should be determinable from the differential change in magnitude between stars of different spectral types.
In practice, we found that PWV measurements from the dual-band GPS system at KPNO did a significantly better job of predicting $z$-band zero-point variation than using the differential change in brightness of stars of different spectral types.
Further more, the dual-band GPS-measured PWV allowed us to successfully correct the observed stellar-color-dependent errors along with the explaining the non-secular zero-point variations.

Despite its non-smooth nature, PWV absorption can be reasonably well-accounted for in a simple linear treatment for SEDs that are smooth.
Specifically, the difference is very small for O--K stars because the SEDs of these stars are well-approximated as smooth blackbodies and $z$-band is on the Rayleigh-Jeans self-similar tail for these objects.
However, the difference is noticeable for M dwarfs and more so for even cooler stars both because the SED becomes non-monotonic within the $z$ band pass as the peak shifts through the band pass and because as molecules start to form in the stellar atmosphere the SED deviates noticeably from that of a blackbody.
This difference has more significant consequences when observing objects with non-stellar SEDs across of range of redshifts, such as supernovae and quasars.
In addition, for time-domain science with strict requirements on accuracy and precision, such as SN~Ia cosmology or exoplanet characterization around M dwarfs, a dual-band GPS system can provide more precise measurements of PWV on a per-observation basis.

An ideal calibration system would utilize all four of the following complementary methods:
(1) large-scale forward-modeling of repeated visits;
(2) narrow-band imaging of stars;
(3) contemporaneous stellar spectra;
and (4) dual-band GPS measurements.
Reliance on a single method is high-risk and will result in uncalibrated data with a method fails.
Relying on just two methods means you use one to calibrate and the second to check; what do you do when they disagree?
Having three or more different methods available allows for a more robust determination with the ability to identify aberrations in any one particular method.

The FGCM model used in DES Y3A1 calibration successfully took advantage of \#1 and \#4.
Future DES calibrations should be able to add in data from aTmCAM~\citep{Li14} and provide a comparison of the two different auxiliary methods of determining PWV.

The list is not precisely parallel.
While \#2--4 are distinct methods with different instrumentation, \#1 is an analysis approach: a forward modeling approach can be used not just for the main survey science data, but could also make use of information from \#2--4 by including those results as data that the forward model could either incorporate or seek to explain.

We recommend that astronomical observatories
install and keep dual-band GPS monitoring systems at observatories.
Large surveys may eventually have enough repeated measurements to calibrate out PWV absorption for the stellar population, and with a forward-modeling approach potentially for more general SEDs.
But even the calibration of the DES Y3 data, which had ample multi-epoch and multi-filter data, was clearly improved by using independently measured PWV from auxiliary sources.
Smaller programs, or single-pass surveys, are less able to make such corrections.
A dual-band GPS system is inexpensive and provides specific and accurate information about the PWV impact on atmospheric transmission.

\acknowledgments

Our thanks to Dustin Lang, David Schegel, Arjun Dey and the DECALS team for rapid and helpful responses to questions we had in the analysis of this data.
Part of the catalyst for this paper was a conversation between M.W.-V. and D. Schlegel at the 20th Anniversary Subaru Meeting hosted by the NOAJ in Waikola, Hawai'i.

This work was supported in part by the US Department of Energy Office of Science under DE-SC0007914.

The Legacy Surveys consist of three individual and complementary projects: the Dark Energy Camera Legacy Survey (DECaLS; NOAO Proposal ID \# 2014B-0404; PIs: David Schlegel and Arjun Dey), the Beijing-Arizona Sky Survey (BASS; NOAO Proposal ID \# 2015A-0801; PIs: Zhou Xu and Xiaohui Fan), and the Mayall z-band Legacy Survey (MzLS; NOAO Proposal ID \# 2016A-0453; PI: Arjun Dey). DECaLS, BASS and MzLS together include data obtained, respectively, at the Blanco telescope, Cerro Tololo Inter-American Observatory, National Optical Astronomy Observatory (NOAO); the Bok telescope, Steward Observatory, University of Arizona; and the Mayall telescope, Kitt Peak National Observatory, NOAO. The Legacy Surveys project is honored to be permitted to conduct astronomical research on Iolkam Du'ag (Kitt Peak), a mountain with particular significance to the Tohono O'odham Nation.

This work is based in part on observations taken at Kitt Peak National Observatory, National Optical Astronomy Observatory, which is operated by the Association of
Universities for Research in Astronomy (AURA) under a cooperative agreement with the National Science Foundation.

\appendix
\section{CCD Temperature Variation Was Not a Source of Significant Zero Point Variation}

CCD temperatures can have a significant effect on the $z$-band QE.  See \citet{Groom17} for a detailed discussion of the transmissions for these thick CCD sensors used in Mosaic-3.
As we approach the Si band-gap energy, single- and double-phonon-assisted conversion of photons becomes at first a significant and then the dominant source of promotion of electrons into the conductance band.
We thus additionally look at CCD temperatures for the $z$-band variations from MzLS.

\begin{figure}
\plotone{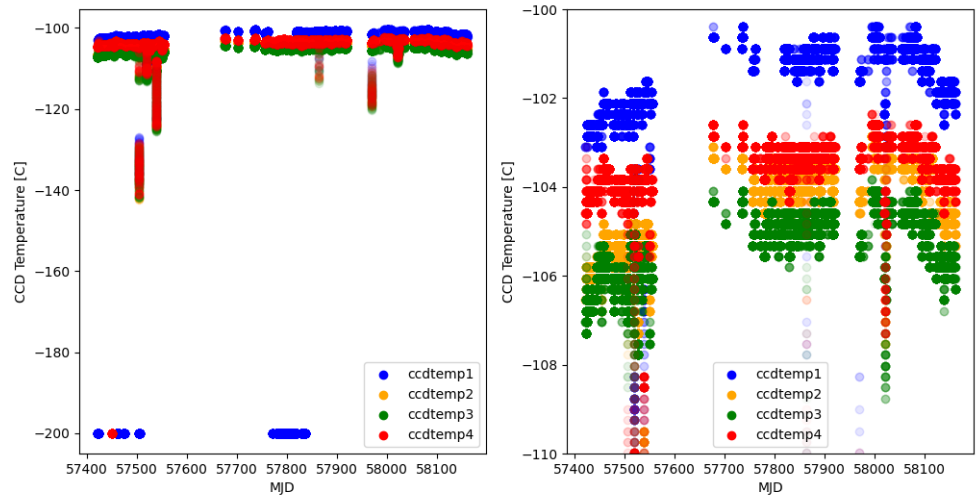}
\caption{
The CCD temperatures over the two years of the survey.
Variation was about 2~C for each CCD, and each CCD had an $\sim$1~C offset with respect to the other CCDs.
(left) full temperature range -- note the 6 outlier nights (the first two nights appear together on this plot).
(right) zoom on the typical range during the survey.
The discreteness in the temperature measurements is due to a 0.246~C step size in the digitization of the signal.
We interpret the recorded values at -200~C as anomalous.
}
\label{fig:ccdtemp_mjd}
\end{figure}

Figure~\ref{fig:ccdtemp_mjd} shows the CCD temperatures for the Mosaic-3 camera over the course of the MzLS survey.
Each CCD stayed within 2~C of its typical temperature over the course of the the MzLS survey (see Table~\ref{tab:ccd_temp}), with the exception of 6 outlier nights.
We compiled these data by extracting the CCDTEMP{1,2,3,4} values from the FITS file headers of the files on disk at NERSC.

\begin{deluxetable}{rr}
\tablecaption{MzLS CCD Median Temperature
\label{tab:ccd_temp}}
\tablehead{\colhead{CCD} & \colhead{Median temperature [C]}}
\startdata
1   & -101.375 \\
2   & -104.332 \\
3   & -105.316 \\
4   & -103.593 \\
\enddata
\end{deluxetable}

There are six nights in the survey where the CCD temperatures were significant outliers.  See Table~\ref{tab:ccd_temp_outliers}.
There is one night when the CCD temperatures were at $< -125$~C.
There are five more nights when the CCD temperatures were at $< -110$~C.
There is no significant zero-point dependency detectable on these nights, it does not
noticeably affect the variance of (MJD, PWV)-corrected ZPT.

\begin{deluxetable}{lrr}
\tablecaption{MzLS CCD Temperature Outlier Nights
\label{tab:ccd_temp_outliers}}
\tablehead{\colhead{Observing Night [MST]\tablenotemark{a}} & \colhead{MJD} & \colhead{T [C]}}
\startdata
2016-04-25 & 57504 & -125 -- -142 \\
2016-04-26 & 57505 & -110 -- -112 \\
2016-05-10 & 57519 & -110 -- -125 \\
2016-05-30 & 57539 & -110 -- -112 \\
2017-04-20 & 57864 & -110 -- -112 \\
2017-08-03 & 57969 & -110 -- -120 \\
\enddata
\tablenotetext{a}{UTC is 6 hours ahead of MST, and the Observing Night is defined as the local date at the start of the evening, so the MJD above is for the next UTC date.}
\end{deluxetable}

\bibliographystyle{aasjournal}
\bibliography{mzls_pwv}

\end{document}